\documentstyle[psfig,preprint,aps]{revtex}
\tightenlines
\def\Journal#1#2#3#4{{#1} {\bf #2}, #3 (#4)}
\def\ADR{{\em Ann. der Phys.}}

\def\CMP{{\em Comm. Math. Phys.}}
\def\JAP{{\em J. Appl. Phys.}}

\def\JPS{{\em J. Phys. Soc. Jpn.}}

\def\NPB{{\em Nucl. Phys.} B}
\def\NPBPS{{\em Nucl. Phys.} B Proc. Suppl.}
\def\PHA{{\em Physica} A}

\def\PLA{{\em Phys. Lett.} A}
\def\PLB{{\em Phys. Lett.} B}
\def\PRL{\em Phys. Rev. Lett.}
\def\PR{{\em Phys. Rev.}}

\def\PRB{{\em Phys. Rev.} B}
\def\PRD{{\em Phys. Rev.} D}

\def\PRSLA{{\em Proc. Roy. Soc. London} A}
\def\PZS{{\em Phys. Z. Sowjet}}

\def\RPP{{\em Rep. Prog. Phys.}}

\newcommand{\be}{\begin{equation}}
\newcommand{\ee}{\end{equation}}
\newcommand{\bea}{\begin{eqnarray}}
\newcommand{\eea}{\end{eqnarray}}

\newcommand{\hf} {{1\over2}}
\newcommand{\nonu}{\nonumber\\}
\newcommand{\br}  {\hskip -0.25cm /}

\def\eq#1{(\ref{#1})}
\def\mathrm{}
\def\tr{{\mathrm tr}}
\def\ph{\hat p}
\def\cp{{\cal P}}
\def\cK{{\cal K}}
\def\cM{{\cal M}}
\def\cb{{\cal B}}
\def\tp{\tilde p}
\def\ra{\rangle}
\def\la{\langle}

\begin{document}
\title{Periodic vacuum and particles in two dimensions}
\author{Marianne Dufour Fournier$^a$\thanks{Marianne.Dufour@IReS.in2p3.fr},
Janos Polonyi$^{ab}$\thanks{polonyi@fresnel.u-strasbg.fr}}
\address{$^a$Laboratory of Theoretical Physics, Louis Pasteur University\\
3 rue de l'Universit\'e 67087 Strasbourg, Cedex, France\\}
\address{$^b$Department of Atomic Physics, L. E\"otv\"os University\\
Puskin u. 5-7 1088 Budapest, Hungary}
\date{\today}
\maketitle

\begin{abstract}
Different dynamical symmetry breaking patterns are explored for the 
two dimensional $\phi^4$ model with higher order derivative terms. 
The one-loop saddle point expansion predicts
a rather involved phase structure and a new Gaussian critical line. 
This vacuum structure is corroborated by the Monte Carlo method, as well.
Analogies with the structure of solids, the density wave phases
and the effects of the quenched impurities are mentioned.
The unitarity of the time evolution operator in real time is
established by means of the reflection positivity.
\end{abstract}

\section{Introduction}
The condensates occuring in Quantum Field Theory are usually homogeneous
and are composed of particles with vanishing momentum. In this manner
the momentum of the excitations is conserved even when a particle is borrowed
from or lended to the vacuum. But the momentum conservation observed
experimentally at finite energies is actually not incompatible with certain 
inhomogeneous vacuua so long as the momentum of the condensed particles is
beyond the observational range. The elementary excitations in solids are 
described by the Bloch 
waves which can be rearranged into different sub-Brillouin zones
in such a manner that the Bloch momentum, the momentum counted from
the center of the sub-Brillouin zone, is conserved. More formally,
the presence of a crystalline ground state restricts the translations 
as symmetries such that the
primitive unit cells are mapped into each other. The conserved quantum
number due to such a restricted symmetry group is the Bloch momentum.
The umklapp processes which take place at the length scale of 
the primitive unit cell change the sub-Brilllouin zone and can be 
interpreted as a change of the type of the excitations. 
Returning now to Quantum Field Theory, one might send
the size of the primitive unit cell to zero. If this
is possible then the
space-time structure of the momentum non-conserving umklapp process 
is not resolved by finite measurements and their interpretation
as a flavor changing process becomes compatible with the experiments.

In order to gain more insight into the role the inhomogeneity of the
vacuum plays in forming the dynamics of the excitations we consider
a generic model with higher order derivative terms in the action
for a scalar field in two dimensions and present its
phase structure, the excitation spectrum and the particle content
when the vacuum possesses a modulation and becomes inhomogeneous.
We use the saddle point approximation in the one-loop order in the analytical
computation and the Monte Carlo for the numerical simulation to have a more
complete picture. The higher order derivatives lead in general to 
the appearance of additional particles with negative norm and 
complex energy. It is pointed
out that our model possesses the reflection positivity which in
turn assures the existence of the positive norm Hilbert space and 
the unitarity of the time evolution operator in Minkowski space-time.

The organization of the paper is the following. Section II contains
our motivation in choosing the model investigated. The tree level vacuum
is identified in Section III. The action is rewritten in terms of the
Bloch waves corresponding to the different periodic vacuua in
Section IV. We diagonalize the quadratic part of the action and 
determine the elementary excitations for the simplest inhomogeneous 
vacuum in Section V. Section VI contains the
demonstration of the one-loop renormalizability of our model.  
The analytical results are compared with a Monte Carlo 
computation in Section VII. The issue of the unitarity 
is discussed in Section VIII. 
Finally Section IX is for the conclusions.

\section{The model}
Our model is an extension of the Landau-Ginzburg model
for a scalar order parameter by adding higher order derivatives
to the action,
\be
S[\phi(x)]=\int d^d x\biggl\{
\hf\partial_\mu\phi(x)\cK\biggl({(2\pi)^2\over\Lambda^2}
\Box\biggr)\partial_\mu\phi(x)+V(\phi(x))\biggr\},\label{lagr}
\ee
where the kinetic energy contains the functions
\be
\cK(z)=1+c_2z+c_4z^2,~~~V(\phi)={m^2\over2}\phi^2+{\lambda\over4}\phi^4,
\ee
and $\Lambda$ is the ultraviolet cutoff.

As far as the dimensionless parameters $c_2$ and $c_4$ are
concerned, we have two different motivations for their introduction.
One is based on the fact that we are always confronted in Nature with effective
theories where the high energy particle exchanges generate a number of
operators in the action which are perturbatively non-renormalizable.
As an example, consider a renormalizable model for a heavy and light particle, 
described by the fields $\Phi$ and $\phi$, respectively and the bare action
$S_0[\phi,\Phi]$. The effective action for
the asymptotic states below the threshold of the heavy particle 
is given by $S_{eff}[\phi]=S_0[\phi,\Phi=0]+\Delta S[\phi]$, where
$\Delta S[\phi]$ contains all effective vertices generated by the heavy
particle exchange processes. In this manner we can never be sure that the
action corresponding to the interactions in a given energy range is actually
limited by the renormalizability even thought the "Theory of Everything"
is supposed to be finite or renormalizable. The higher order
derivative terms of our model may arise from $\Delta S[\phi]$.
The decoupling theorem, \cite{ac}, helps us out from the problem of
a too general action with nonrenormalizable terms by
asserting that the nonrenormalizable coupling constants are small, being
suppressed by the power of the light and the heavy particle mass ratio. 
The model \eq{lagr}, a slight extension of an effective theory
for the Higgs boson \cite{hiderk}, retains some of the suppressed
nonrenormalizable terms arising from a hitherto unknown 
super heavy particle exchange.

The question, left open by the conclusion of the decoupling theorem,
and which motivates the present work is whether the smallness of the
non-renormalizable coupling constants is really sufficient to render
them unimportant in the effective theory. We shall find that certain 
higher order derivative terms may become relevant when their coupling
constants exceed a threshold value. In other words, we might be
forced to consider non-renormalizable terms in our effective theories
if the heavy particle is not exceedingly far from the observational energy. 
The other motivation to study the model \eq{lagr} with non-renormalizable
terms is the suspicion that the perturbatively
non-renormalizable terms might turn out to be relevant by a non-perturbative
mechanism and actually allow the removal of the cutoff. If this is happened 
to be the case then our model with non-vanishing $c_2$ and $c_4$ is as 
justified as the usual one whose action is quadratic in the gradient,
a possibility motivating the authors of ref. \cite{hiderk}.

What kind of heavy particle exchange is behind the higher order derivative
terms of our action? 
As mentioned above, our interest is in theories with inhomogeneous
vacuum. Such a ground state which is modulated with a period length $\lambda_{vac}$ 
is the result of a force which is attractive for $x>\lambda_{vac}$
and repulsive when $x<\lambda_{vac}$. 
When the particle whose exchange generates this force 
is eliminated, its effects are kept in the choice of the vertices in 
$\Delta S[\phi]$. The momentum independent ultra-local (i.e. 
non-derivative) terms which contribute
to the local potential in the action can not generate such a strong
distance dependence in the interaction. But it is easy to see that the 
higher order terms in the derivative are just for this role, to
lower the action for modes whose characteristic momentum scale is
$p\approx1/\lambda_{vac}$. In fact, consider the eigenvalues 
of the second functional derivative
of the action evaluated in the trivial vacuum $<\phi(x)>=0$,
\be
\epsilon(p)=V''(0)+p^2\cK\left(-{(2\pi)^2\over\Lambda^2}p^2\right).
\label{inprop}
\ee
The $O(p^4)$ term produces a non-trivial local minimum
at $p=p_{min}\approx\lambda_{vac}^{-1}$ for $c_2,~c_4>0$. Thus $c_2$ and
$c_4$ correspond to a van der Waals force.
We shall go further in this work and ask what happens when $c_2$ 
reaches so large values that the minimum of the dispersion relation
$\epsilon(p)$ turns out to be negative.
The corresponding vacuum will be the condensate of particles with 
momentum around $p_{min}$ and will display the period length
$\lambda_{vac}\approx p_{min}^{-1}$. 

The hand-waving argument to retain the higher order derivative terms
of \eq{lagr} from the multitude of other contributions in $\Delta S[\phi]$ is the
following. Let us start with $c_4=0$, when the vacuum is homogeneous 
for $c_2<0$. On the contrary, for $c_2>0$ an instability opens by increasing 
the momentum of the condensed particles and the
ultraviolet cutoff stabilizes the vacuum \cite{riv} where particles with
momentum at the cutoff are found. Models with such an instability
were studied in refs. \cite{zarh}, \cite{zarn} and \cite{ehj} in three and 
four dimensions. If the inhomogeneous vacuum is
supposed to be formed at momentum scales below the cutoff then
we need another
stabilization mechanism. For this end we retain the $O(\partial^6)$ term
with $c_4>0$. We believe that the higher
order terms in the gradient will modify the shape of the saddle point
only leaving the qualitative features of the inhomogeneous vacuum
unchanged. In other words, the kinetic energy is generic. 
The present work can be considered as the continuation of Ref. \cite{ehj}
where a new gaussian ultraviolet fixed point was found in the one-loop
approximation for $c_4=0$. 
We simplify in this work the issue of the renormalizability 
by choosing lower dimension, $d=2$, but the
multitude of different phases is explored by allowing $c_4\not=0$.

The allusion made above at the Landau-Ginzburg model is based on the 
similarity of the functions $V(u)$ and $\cK(u)$. The nontrivial 
absolute minima provide the mechanism of the spontaneous 
($p=0$) or dynamical ($p\not=0$) symmetry breaking generated by 
the potential in the internal space or the kinetic energy in 
the external (and internal) space, respectively. Our interest in 
this work is to explore the different dynamical symmetry breaking patterns
provided by the generic action \eq{lagr} and to suggest a mean-field
treatment for the phase transitions with modulated ground state \cite{str}.

The questions addressed here and their tentative answers
have certain relevance both in Solid State
and High Energy Physics. The nontrivial, periodic vacuum generated
by the higher order derivative terms may offer a new point of view
in understanding the origin of the crystalline structure in solids.
In fact, consider the coupled system of electrons, ions and photons.
It is already an effective theory because the lower lying electrons
of the ions are represented by the insertion of different charge 
distributions and form factors for the ions. We introduce 
chemical potentials for the electrons and the ions in order to
realize electrically neutral matter with finite density. 
Finally we eliminate the heaviest degrees of freedom, the ions.
This is opposite to the usual Born-Oppenheimer approximation
but it generates a local effective interaction. This effective
theory for electrons and photons contains higher order derivative 
terms for the photon field which yields the periodic,
crystalline ground state. Another appearances of this mechanism 
where strong van der Waals forces are acting are
the antiferromagnets and 
the charge or density wave phases. In the latter the effective theory is
obtained by eliminating the valence electrons and has periodic
tree level ground state \cite{cdw}. 
The massless case with $\epsilon(p_{min})>0$ may as well represent a superfluid system
with $p_{min}$ as the roton momentum \cite{land}. In this context our
quantitative argument showing the relation between the van der Waals
forces and the higher order derivatives in the action represents a
simple, effective theory motivated alternative of Feynman's argument about
the rotons \cite{feynm}. The instability leading to the formation of
an inhomogeneous ground state when $\epsilon(p_{min})<0$ is a
quantum phase transition where a non-classical soft mode \cite{qpt} shows up
at $p\approx p_{min}$. 

The periodic vacuum of our model supports frustrations for certain choice
of the coupling constants. To see this we write the lattice regularized
version of \eq{lagr} in the form
\bea\label{res}
S[\phi(x)]&=&
\sum_x\biggl\{-\hf\phi(x)\Box\cK(\Box)\phi(x)+V(\phi(x))\biggr\}\nonu
&=&\sum_x\biggl\{\phi(x)\biggl[
A\phi(x)+\sum_\mu\biggl(B(\phi(x+e_\mu)+\phi(x-e_\mu))\nonu
&+&C(\phi(x+2e_\mu)+\phi(x-2e_\mu))
+D(\phi(x+3e_\mu)+\phi(x-3e_\mu))\biggr)\nonu
&+&\sum_{\mu\not=\nu}\biggl(E(\phi(x+e_\mu+e_\nu)+\phi(x+e_\mu-e_\nu)
+\phi(x+e_\nu-e_\mu)+\phi(x-e_\mu-e_\nu))\nonu
&+&F(\phi(x+2e_\mu+e_\nu)+\phi(x+2e_\mu-e_\nu)+\phi(x-2e_\mu+e_\nu)\nonu
&+&\phi(x-2e_\mu-e_\nu))\biggr)\\
&+&G\sum_{\mu\ne\nu\ne\rho}\biggl(\phi(x+e_\mu+e_\nu+e_\rho)
+3\phi(x+e_\mu+e_\nu-e_\rho)\nonu
&+&3\phi(x+e_\mu-e_\nu-e_\rho)+\phi(x-e_\mu-e_\nu-e_\rho)\biggr)
\biggr]+{\lambda\over4}\phi^4(x)\biggr\},\nonumber
\eea
where $A=m^2/2+d-(2d^2+d)c_2+(4d^3+6d^2)c_4$, $B=-1/2+2dc_2-(6d^2+3d-3/2)c_4$,
$C=E/2=-c_2/2+3dc_4$ and $D=F/3=G=-c_4/2$. The possibility of having either
sign for these coefficients indicates the competition between the 
nearest- and beyond nearest-neighbor interactions and the possible
presence of frustrations as lattice defects of the periodic vacuum. 
When the amplitude of the periodic vacuum is large the motion of the
frustration is rather slow and the model offers a semiclassical 
description of the quenched disorder.

The usual strategy in High Energy Physics is to introduce a field
variable for each particles. But models with inhomogeneous vacuua may
display more involved particle-field assignments by exploiting
the non-trivial dynamics at the ultraviolet cutoff scale. In fact, there
are several dispersion relations and particle like-elementary excitations
in Solid State Physics, such as the acoustical and the optical phonons
and the massless or massive excitations in antiferromagnets. In both
examples the dynamics is rather non-trivial at the ultraviolet cutoff.
Can we keep such a more unified description of several particles
by means of a single quantum field in a renormalizable model? In this case
the dynamics at the cutoff can be pushed at infinitely high energies and
it is not obvious that the construction converges. Our answer to this
question is affirmative up to the one-loop order of the perturbation expansion. 
Another interesting
aspect of the model considered is the possibility of breaking continuous
external symmetries without generating massless Goldstone modes. In fact,
suppose that we use say lattice regulator which breaks the continuous translation
invariance and the momentum of the condensed particles is close to the
cutoff, in which case the continuum description is not applicable
at the length scale of the vacuum and there is no reason to expect
gapless excitations. This is another manifestation of the apparent 
homogeneity of the modulated vacuum with shrinking period length.

There are several works devoted to the applications of 
models with higher order derivatives. 
The canonical formalism for Lagrangians with
higher order derivatives has been worked out in ref. \cite{ostr} and
been applied to QED \cite{pauh}, \cite{lewi}, gravity \cite{hiderg}, 
the Higgs sector of the Standard Model \cite{hiderk} and Chern-Simons
theories \cite{hidercs}. The present paper is close in spirit and the method
followed to the  nonperturbative lattice study of the effects of the higher 
derivative terms for a scalar particle \cite{hiderk}. But contrary
to that paper where the continuum limit is sought in the 
symmetrical and the ferromagnetic phase by keeping both the particle mass
and the scale induced by the higher order derivative terms finite
we shall consider the phase with inhomogeneous vacuum whose length scale does
shrink to zero with the cutoff.

The higher order derivative terms may
generate new tricritical point, called the Lifshitz point. 
One can see this in the Euclidean field theory given by the lagrangian
\be\label{lipol}
L={c_1\over2}(\partial_\mu\phi)^2+{c_2\over2}(\Box\phi)^4
+{g_2\over2}\phi^2+{g_4\over4!}\phi^4+{g_6\over6!}\phi^6.
\ee
It is well known that the model $c_2=0$, $g_6>0$ possesses a tricritical point 
which is at $g_2=g_4=0$ in the mean-field approximation.
For $g_4>0$ the phase transition following the
sign of $g_2$ is changed is of second order. When $g_4<0$
then the crossing of the line $g_2=5g_4^2/8g_6$ corresponds
to a first order phase transition whose latent heat vanishes 
as $g_4\to0^-$. A similar tricritical
point is generated by the competing derivative terms for $g_6=0$
at $g_2=c_1=0$, according to an $\epsilon$-expansion like 
renormalization group method \cite{lipo} and the solution of 
the spherical model \cite{lipos}.
Phase transitions of different type coincide here.
The change of the sign of $g_2$  
induces a second order phase transition when $c_1\ge0$.
$\la\phi(x)\ra$ is inhomogeneous for $c_1<0$, and
its wave vector approaches zero 
as $c_1\to0^-$. All of these studies deal with the
quantum or statistical physics of the fluctuations around the 
trivial vacuum, $\la\phi(x)\ra=0$. We go beyond this level by allowing
the condensation in the ground state and considering
the fluctuations around a periodically modulated
mean field $\la\phi(x)\ra$.

The interference of the cutoff and the derivative terms leads to
commensurate-incommensurate transitions \cite{incom}, as well.
The simplest 1+1 dimensional model of these transitions 
is based on the lattice potential energy \cite{frko}
\be\label{frko}
U=\sum_x\left[{z\over2}(\phi(x+1)-\phi(x)+\delta)^2
-{g\over2}\cos{2\pi\phi(x)\over b}\right].
\ee
For large $g$ the value of the field tends to take integer multiples
of $b$, the period length of the second term in the potential.
When $z$ is chosen to be large then the average increment
of the field at neighboring sites, $\bar\delta$, is close to $\delta$. The 
configurations which minimize the energy are governed by
the competition between the conflicting requirements of the
two contributions in \eq{frko}.
The ratio $\bar\delta/b$ plotted against $\delta/b$
exhibits a characteristic, discontinuous structure, called the devil's
staircase. Similar competition is expected between the
period length of the vacum $\lambda_{vac}$ introduced at eq.
\eq{inprop} and the lattice spacing in the solution of the
equation of motion of \eq{res}. Our model offers
a translation invariant realization of the
commensurate-incommensurate transitions. Furthermore the sign of 
$\partial_\mu\la\phi(x)\ra$ is given by
the dynamical breakdown of the space inversion symmetry
instead by the explicit symmetry breaking term, $\delta$, in \eq{frko}.

We close this section by mentioning a formal problem which poses a serious
threat in using actions with higher derivative terms \cite{pauh}. The inverse
propagator \eq{inprop} has several roots as the function of $p^2$.
Each of them yields a (not necessarily) simple pole for the energy
integrals which is passed by the integration contour during
the Wick rotation of the perturbation series
into Minkowski space-time. When the real part of the Euclidean energy at a simple
pole is non-vanishing then the Wick rotation produces a particle mode with
complex energy and with amplitude which growths or decreases 
exponentially in time. Even if
the Euclidean theory is stable we have to make sure that such runaway
amplitudes are canceled after the Wick rotation \cite{lewi}, \cite{cula}. 
Since the original theory, the source of our effective model 
was supposed to have stable vacuum in the Minkowski space-time 
this problem may only arise from the inconsistent truncation of the effective 
lagrangian \eq{lagr}. 

One encounters a similar problem in the Schwinger-Dyson
resummation procedure, as well. The effective action which approximates
the Wilsonian renormalized action in the infrared limit, as the cutoff 
tends to zero, has a non-trivial momentum dependent piece which is quadratic in the
field, the self energy. When the self energy is considered in a
non-perturbative manner then its momentum dependence may create new poles.
These are usually ignored since they imply that the self energy, coming from
the radiative corrections is as important as the original $p^2$ kinetic energy
term. But their proper treatment leads to the question of the consistence of the
truncation of the effective action.

This point of view opens a general question.
All theory we know in Physics is effective only and contains
higher order derivatives, whatever small coefficients they may have.
How can we make sure that the studies restricted to the models with 
the usual kinetic energy reflect what happens in 
the effective theories?
The answer to this question involves the verification of the
usual universality scenario in the mixed framework of the
static and the dynamical renormalization group, where the 
fixed point is in the ultraviolet for the spatial momenta and
in the infrared for the energy. Instead of embarking this
important but
extremely involved issue we restrict ourselves in this work
to show that our model is consistent in real time
and the runaway modes can be excluded in a nonperturbative manner.

\section{Tree level vacuum}\label{mfsol}
We start the semiclassical solution of our model by determining the
minimum of the lattice action. This is in principle a rather involved numerical
problem, the minimization of \eq{res}. To circumvent this complication
we shall seek the tree level, mean-field vacuum in the form 
\be
\phi_{MF}(x)=\phi_H+\phi_{IH}\cos\left(K\sum\limits_{\mu=1}^{d_{AF}}
x^\mu+\theta\right),~~~
K=2\pi{M\over N}\label{trconf}
\ee
where the amplitudes $\phi_H$, $\phi_{IH}$, the relative primes 
$N$, $1\le M\le N/2$,
the phase $\theta$ and the number of the antiferromagnetic 
directions, $d_{AF}=1,2$ serve as the variational parameters to minimize 
the action. We have naturally to confine our study into regions far from the critical
points, i.e. above the Ginzburg temperature in order to apply this method.
The phase is called para-, ferro-, antiferro- and
ferri-magnetic for $\phi_H=0=\phi_{IH}$, $\phi_H\not=0=\phi_{IH}$,
$\phi_H=0\not=\phi_{IH}$ and $\phi_H\not=0\not=\phi_{IH}$.
The action density, $s(d_{AF},M,N)=S/L^2$, on a lattice $L\times L$ with $m^2<0$ is $s_0=-m^4/4\lambda$, for $d_{AF}=0$. 

For the computation of the action for $d_{AF}>0$ we restrict ourselves 
to the case $m^2<0$. The mean-field vacuum configuration is an
eigenvector of the lattice box operator,
\be
\Box\phi_{MF}(x)=-4d_{AF}\sin^2\left({K\over2}\right)(\phi_{MF}(x)-\phi_H),
\ee
what gives
\be
-\Box\cK(\Box)\phi_{MF}(x)=\cM^2(M,N,d_{AF},c_2,c_4)(\phi_{MF}(x)-\phi_H),
\ee
with
\be
\cM^2=4d_{AF}\sin^2\left({K\over2}\right)
\left[1-4d_{AF}\sin^2\left({K\over2}\right)c_2
+16d^2_{AF}\sin^4\left({K\over2}\right)c_4\right].\label{propg}
\ee
The dependence of the eigenvalue $\cM$ in the parameters $M,N,d_{AF},c_2,c_4$
will be suppressed in the expressions below. The action density to minimize is
\bea
s&=&{m^2\over2}\phi_H^2+{C_2(N)\over2}(m^2+\cM^2)\phi_{IH}^2\nonu
&&+\lambda\biggl({1\over4}\phi_H^4+{3C_3(N)\over4}\phi_H\phi_{IH}^3
+{3C_2(N)\over2}\phi_H^2\phi_{IH}^2
+{C_4(N)\over4}\phi_{IH}^4\biggr),\label{varpo}
\eea
where we introduced the notation 
\be
C_n(N)=\left[{1\over N}\sum\limits_{\ell=1}^N\cos^n
\left(2\pi{M\ell\over N}+\theta\right)\right]^{d_{AF}}.
\ee
Notice the $M$-independence of the sum for the relative primes $M,N$.
Direct computation gives
\bea\label{cs}
C_2(N)&=&\cases{\left({1+\cos2\theta\over2}\right)^{d_{AF}}&$N=2$,\cr
2^{-d_{AF}}&$N>2$,}\nonu
C_3(N)&=&\cases{\left({\cos3\theta\over4}\right)^{d_{AF}}&$N=3$,\cr
0&$N\not=3$,}\\
C_4(N)&=&\cases{\left({3+4\cos2\theta+\cos4\theta\over8}\right)^{d_{AF}}
&$N=2$,\cr
\left({3+\cos4\theta\over8}\right)^{d_{AF}}&$N=4$,\cr
\left({3\over8}\right)^{d_{AF}}&$N=3$ or $N>4$.}\nonumber
\eea
The dependence of $C_n$ in $N$ will not be shown explicitly below.
One can see from the above construction that the mean-field solution of
a model $\phi^\ell$ requires the coefficients $C_n$ with 
$n=1,\cdots,\ell$. So the limit $N\to\infty$,
the regular dependence on $N$ sets on for $N>\ell$.

The action density corresponding to different choices of $N$
is obtained as follows:

\underbar{$N=2$}: The phase parameter $\theta$ is redundant in this case.
We choose $\theta=0$, what sets $C_2=C_4=1$, and write
\be
s=A\cdot X+\hf X\cdot B\cdot X,\label{acevn}
\ee
where
\be
X=\pmatrix{\phi_H^2\cr\phi_{IH}^2},~~~~A=\hf\pmatrix{m^2\cr m^2+\cM^2},~~~~
B={\lambda\over2}\pmatrix{1&3\cr 3& 1}.
\ee
By the help of the shift $Y=X+X_0$ where
\be
X_0={1\over8\lambda}\pmatrix{-2m^2-3\cM^2\cr -2m^2+\cM^2}.
\ee
we obtain $s=\hf Y\cdot B\cdot Y+s_m$. The rotation matrix
\be
{\cal R}(\Theta)=\pmatrix{\cos\Theta&\sin\Theta\cr-\sin\Theta&\cos\Theta}\label{rot}
\ee
with $\Theta=\pi/4$ diagonalizes the quadratic form,
\be
s=-{\lambda\over4}(Y^1-Y^2)^2+{\lambda\over2}(Y^1+Y^2)^2+s_m.
\ee
The minimum is the result of the competition between the negative and
the positive eigenvalue, i.e. the trend to increase $|Y^1-Y^2|$ and to
decrease $|Y^1+Y^2|$. The result is that the minimum is reached 
at the boundary of the quadrant $X^1\ge0$, $X^2\ge0$, i.e.
there is no ferrimagnetic phase realized. Thus the
mean-field vacuum is found at the minimum of the following
two functions
\be
s(\phi_H^2,0)=\hf m^2\phi_H^2+{\lambda\over4}\phi_H^4,~~~
s(0,\phi_{IH}^2)=\hf C_2(m^2+\cM^2)\phi_{IH}^2+{\lambda\over4}C_4\phi_{IH}^4.
\label{actden}
\ee
The vacuum is antiferromagnetic when
\be
{C_2^2\over C_4}\left(1+{\cM^2\over m^2}\right)^2>1\label{condaf}
\ee
with the action density
\be
s=s_0{C_2^2\over C_4}\left(1+{\cM^2\over m^2}\right)^2.\label{actaf}
\ee

\underbar{$N=3$}: The minimization of the action with respect $\theta$ yields
the condition $\phi_H\phi_{IH}^3\sin3\theta=0$,
whose solutions, $\phi_H=0$, $\phi_{IH}=0$ and $\theta=0, \pi/3$ correspond
to the anti-, ferro- and ferrimagnetic phases, respectively. The transformation
$\theta\to\theta+\pi/3$, and $\phi_H\to-\phi_H$
leaves the mean field action invariant and it is sufficient to consider the 
case $\theta=0$ to explore the ferrimagnetic phase. The action density 
in the ferro- and antiferromagnetic phases is given by \eq{actden}
for $\theta\not=0$. The antiferromagnetic phase is preferred for
\eq{condaf} and the corresponding action density is \eq{actaf}.
For $\theta=0$ we have to minimize \eq{varpo} numerically.

\underbar{$N=4$}: The minimization with respect $\theta$ yields
$\phi_{IH}^4\sin4\theta=0$,
showing the possibility of the ferrimagnetic phase when $\theta=n\pi/4$.
Since the action is an even function of the amplitudes
$\phi_H$ and $\phi_{IH}$ for even $N$ we have the expression \eq{acevn}
where
\be
A=\hf\pmatrix{m^2\cr C_2(m^2+\cM^2)},~~~~
B={\lambda\over2}\pmatrix{1&3C_2\cr 3C_2& 1}
\ee
what results
\be
X_0={1\over\lambda(9C_2^2-C_4)}\pmatrix{m^2(1-3C_2^2)-3C_2^2\cM^2\cr
m^2C_2(C_4-3C_2)+C_2C_4\cM^2}.
\ee
The rotation \eq{rot} satisfying the condition $\cot2\Theta=(1-C_4)/6C_2$ 
transforms the action into the diagonal form with the eigenvalues:
\be
{\lambda\over2}\left(1+C_4\pm\sqrt{(1+C_4)^2+36C_2^2-4C_4}\right),
\ee
indicating that one of the normal modes is again unstable.
Due to the negative eigenvalue and $0<\Theta<\pi/4$, the minimum is always
reached at the boundary of the quadrant $X^1\ge0$, $X^2\ge0$, i.e.
there is no ferrimagnetic phase. 
The condition for the antiferromagnetic phase and the expression
of the action density are given by \eq{condaf} and \eq{actaf}.

\underbar{$N>4$}: The procedure is the same as for $N=4$, with
the only difference is that $C_3=0$ and there is no 
$\theta$-dependence in the sums $C_2$ and $C_4$.

\begin{figure}
\centerline{\psfig{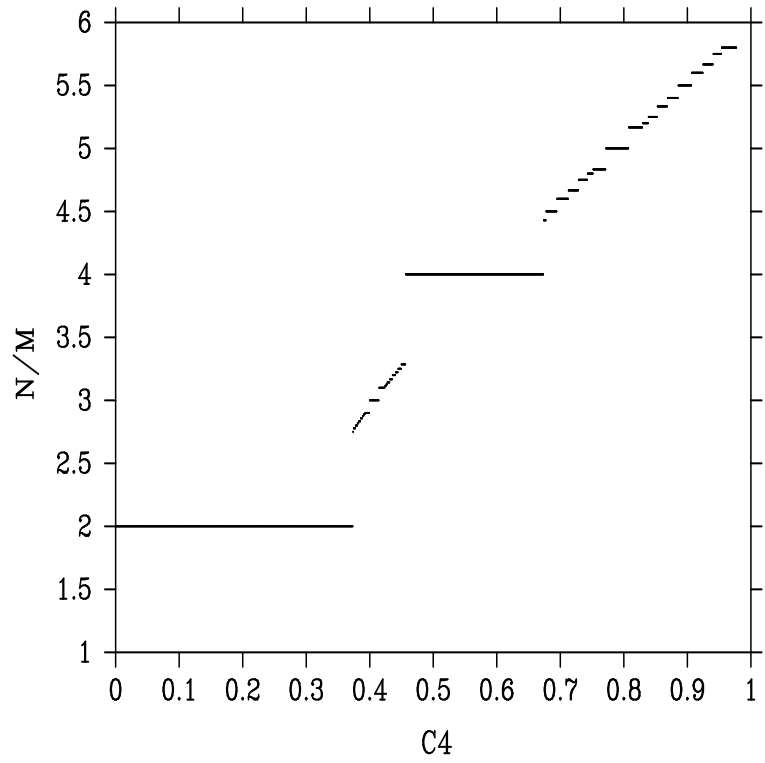}}
\caption{\label{devil} The devil's staircase, the $c_4$ dependence 
of the period length of the vacuum at $c_2=2$.}
\end{figure}

\underbar{Phase structure:}
The resulting action densities are summarized in Table \ref{esperes}.
The $c_4$ dependence of the period length of the vacuum for $c_2=2$ and
$N\le N_{max}=32$
is shown in Fig. \ref{devil}. The general trend is that the increase of
$c_4$ pushes the minimum of the dispersion relation $p_{min}$ towards
zero thereby increasing the period length of the vacuum. At a certain
threshold value the minimum at $p_{min}$ becomes so 
shallow that the potential energy turns the vacuum homogeneous and
the system undergoes a ferromagnetic phase transition. Notice the usual 
signatures of the commensurate-incommensurate transitions, the "devil's
staircase" structure. This is a competition between two length
scales, the cutoff $a$ and the period length of the condensate $\lambda_{vac}$,
\be
M\lambda_{vac}=Na.
\ee
The period length in lattice spacing units, $N(c_2,c_4)/M(c_2,c_4)$ is 
shown in Fig. \ref{devil} as the function of $c_4$. 
It "locks-in", i.e. stays constant in a larger 
commensurate interval where the relative primes $N$ and $M$ are small.
The long strips corresponding to the $N=2$ and 4
phases show a strong "lock-in" effect, contrasted with the gradual
change of the period length for other values of $N$.
The numerator and the denominator as the functions of $c_4$ are non-monotonic
in the same time. 

\begin{figure}
\centerline{\psfig{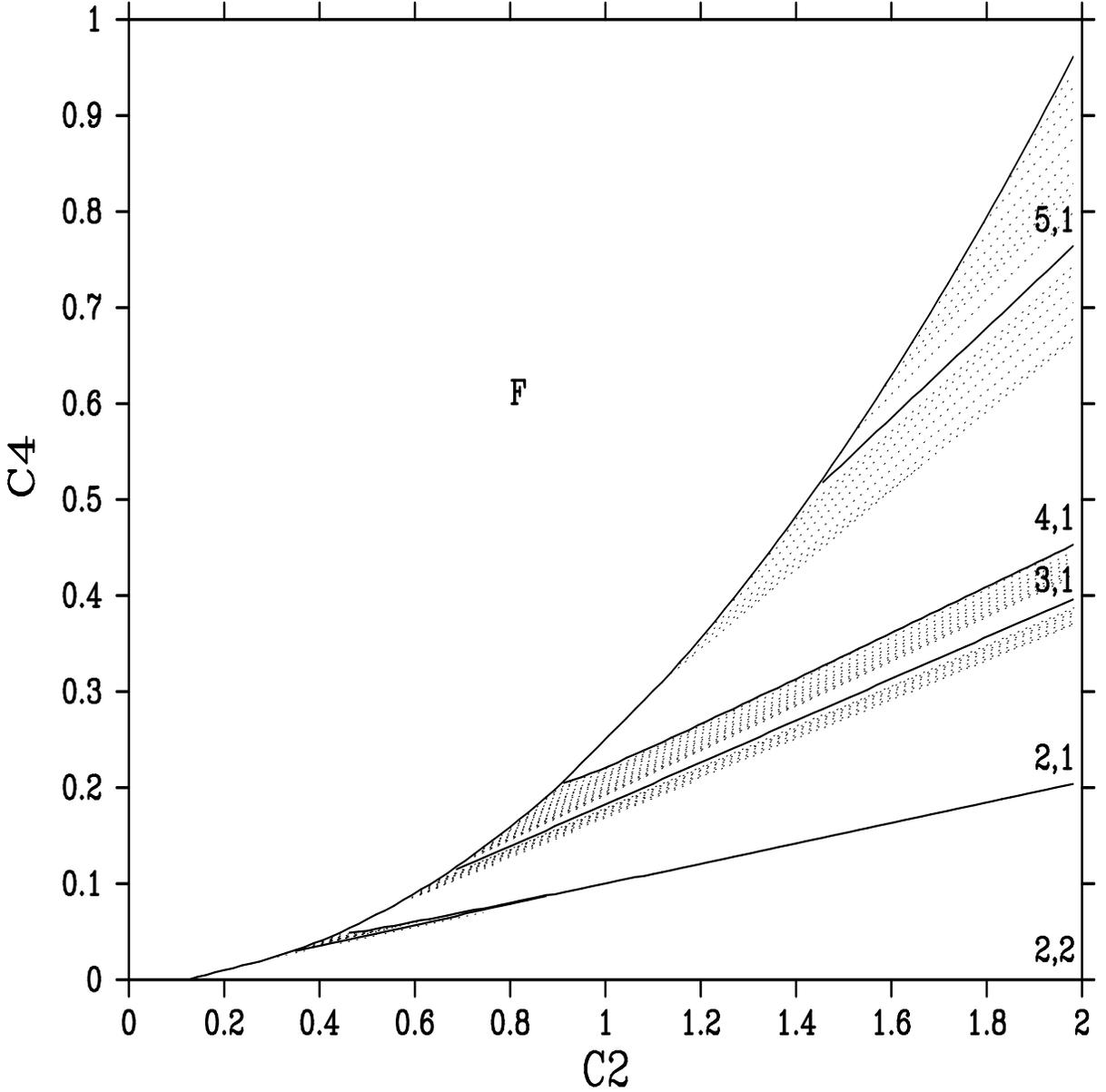}}
\caption{\label{mfphstr} 
The mean field phase structure in the
$(c_2,c_4)$ plane. F denotes the ferromagnetic phase.
The lower edges of the phases with $M=1$ and $M>1$
are indicated by solid and dashed lines, respectively. The values of
$N,d_{AF}$ are shown for $M=1$ at the right.}
\end{figure}

The phase structure in the plane $(c_2,c_4)$ is depicted in Fig. 
\ref{mfphstr}. We identified the mean-field parameters $N\le N_{max}=32$, $M$ and $d_{AF}$
for each point in the plane $(c_2,c_4)$. The points where we enter into 
a phase with $M=1$ by increasing $c_4$ are indicated by the solid lines, 
the other phase boundaries are shown by dotted lines. The increase of 
$N_{max}$ makes the dotted lines denser without changing the solid lines
or populating the white area.
The vacuum in the upper left region is ferromagnetic.
The lower right part of the $c_2,~c_4>0$ quadrant contains the
inhomogeneous vacuua, each of them situated in slightly tilted strips with
increasing $N/M$ as we move upwards. The narrow triangular
phase between the $N=2$ relativistic and non-relativistic phase
is a relativistic phase with $N=3$, see Fig. \ref{mfpr}. It is interesting 
that the relativistic vacuua are realized for $N=2$ and 3 only.
We found no ferrimagnetic phase for the parameters considered.

\section{Excitation bands}
The study of the more detailed structure of the dynamics starts with the 
determination of the possible excitations.
The distinguishing feature of the antiferromagnetic vacuua is 
their inhomogeneity and the nonconservation of the momentum of
the excitations. We introduce in this Section a convenient formalism
for the description of the elementary excitations. Following the solid 
state analogies we rewrite our action by means of the Bloch waves 
which take the non-conservation of the momentum into account.

\underbar{Brillouin zones}:
First we split the first Brillouin zone
\be
\cb=\left\{p_\mu;|p_\mu|\le\pi,~\mu=1,2\right\}
\ee
of the $d$ dimensional $N$ antiferromagnetic phase
into $N^{d_{AF}}$ sub-zones within which the Bloch momentum
is conserved. In the relativistic case when $d_{AF}=d$, we find
\be
\cb^r_\alpha=\left\{p_\mu;|p_\mu-P^{(N)}_\mu(\alpha)|\le{\pi\over N}\right\},
\ee
where the center of the sub-zone, $P^{(N)}_\mu(\alpha)$, is given by 
\be
P^{(N)}_\mu(\alpha)={2\pi\over N}n_\mu(\alpha)\label{pdef}
\ee
in terms of the integer valued vector $n_\mu(\alpha)=0,\cdots,N-1$,
\be
\alpha=1+\sum_\mu n_\mu(\alpha)N^{\mu-1}.\label{value}
\ee
In other words, the integer component vector $n_\mu(\alpha)$ gives the center 
of the sub-zone $\cb_\alpha^r$ in units of $2\pi/N$. In the same time, it can 
be considered as an $N$-base number. In this case its value labels the 
corresponding sub-zones. In the nonrelativistic case, $d_{AF}<d$, we have 
\be
\cb_\alpha^{nr}=\left\{p_\mu;|p_\mu-P^{(N)}_\mu(\alpha)|\le{\pi\over2},~\mu\le d_{AF},~~
|p_\mu|\le\pi,~\mu>d_{AF}\right\}.
\ee

In the next step we introduce a field variable in real space 
which is responsible for the fluctuations in each sub-zone,
\be
\phi_\alpha(x)=\int_{\cb_\alpha}{d^dp\over(2\pi)^d}e^{ip\cdot x}\phi(p)
=\int_{\cb_1}{d^d\tp\over(2\pi)^d}e^{i(\tp+P^{(N)}(\alpha))\cdot x}
\phi_\alpha(\tp)
\ee
by the help of the Fourier transform
\be
\phi(p)={1\over L^d}\sum_xe^{-ix\cdot p}\phi(x),
\ee
and its restriction into the sub-zones, $\phi_\alpha(\tp)=\phi(\tp+P^{(N)}(\alpha))$.
The computation what follows is considerably simplified if the 
$N^{d_{AF}}$ Fourier transforms are extended over the first Brillouin zone 
as periodic functions,
\be
\phi_\alpha(\tp)=\phi_\alpha(\tp+P^{(N)}(\beta)),\label{perft}
\ee
where $\beta$ is an arbitrary sub-zone index. The tilde on a momentum 
variable will always denote that the given momentum is in the sub-zone 
$\cb_1$. The path integral is then written as 
\be
\prod\limits_p\int d\phi(p)e^{-S[\phi]}
=\prod\limits_\alpha\prod\limits_{\tp}\int\phi_\alpha(\tp)e^{-S[\phi_\alpha]},
\ee
with
\bea
L^dS[\phi_\alpha]
&=&\hf\int{d^dp\over(2\pi)^d}\phi(-p)[\cp^2\cK(-\cp^2)+m^2]\phi(p)\nonu
&&+{\lambda\over4}\left(\prod\limits_{k=1}^4\int{d^dp_k\over(2\pi)^d}
\phi(p_k)\right)(2\pi)^d\delta(\sum\limits_kp_k)\label{lagrp}
\eea
where
\be
\cp^2(p)=4\sum_\mu\sin^2{p_\mu\over2}
\ee
denotes the momentum square on the lattice.

\underbar{Flavor algebra}: The $N^{d_{AF}}$ sub-zones introduced above
correspond to the different excitation bands of the coarse grained lattice
whose lattice sites represent the primitive cells of the original lattice.
The umklapp processes where a non-vanishing momentum is exchanged with the
vacuum take the momentum from one sub-zone to another. To separate the
sub-zone preserving and changing processes from each other
we rewrite the momentum integrals
in \eq{lagrp} as a sum over the sub-zones and integration within $\cb_1$,
\be
\int d^dpf(p)=\sum\limits_{\alpha=1}^{N^{d_{AF}}}\int d^d\tp
f(\tp+P^{(N)}(\alpha)).
\ee
The summation can be organized in a more transparent manner by
considering the group 
\be
Z_{d_{AF},N}=\otimes\prod\limits_{j=1}^{d_{AF}}Z_N\label{modmom}
\ee
which describes the shift of the momentum in the periodic, 
$d_{AF}$ dimensional region of the Brillouin zone. This
group makes its appearance by considering the action \eq{lagrp}
as a matrix element of an operator in the function space span by the 
"wave functions" $\phi_\alpha(\tp)$. It is advantageous to use the plane 
wave basis $|P^{(N)}(\alpha)+\tp\ra=|\alpha,\tp\ra$ in which the matrix 
element of the field operator $\phi_\alpha(\tp)$ is defined by
\bea
<\beta',\tilde q'|\phi_\alpha(\tp)|\beta,\tilde q>
&=&\delta\left(
P^{(N)}(\beta)+\tilde q+P^{(N)}(\alpha)+\tp-P^{(N)}(\beta')-\tilde q'
\right)\phi_\alpha(\tp)\nonu
&=&\delta\left(P^{(N)}(\beta)+P^{(N)}(\alpha)-P^{(N)}(\beta')\right)
\delta\left(\tilde q+\tp-\tilde q'\right)\phi_\alpha(\tp)\nonu
&=&(\gamma^\alpha)_{\beta',\beta}
\delta\left(\tilde q+\tp-\tilde q'\right)\phi_\alpha(\tp),
\eea
where the periodicity \eq{perft} was used in the second line.
The symbol $\phi_\alpha(\tp)$ stands for operator when sandwiched
between the basis vectors and for function in the c-number expressions.
We introduced here a representation of $Z_{d_{AF},N}$ by means of 
$N^{d_{AF}}\times N^{d_{AF}}$ matrices, 
\be
\left(\gamma^\alpha\right)_{\rho,\sigma}=\prod_{\mu=1}^{d_{AF}}
\delta_{\sigma_\mu+\alpha_\mu,\rho_\mu(modN)},
\ee
constructed in such a manner that $\gamma^\alpha$ describes the 
effect of the exchange of the momentum $P^{(N)}(\alpha)$ on the index
labeling the sub-Brillouin zones. 
We shall need later a relation arising from the Abelian nature
of the group $Z_{d_{AF},N}$,
\be
(\Gamma)_{\alpha+\rho(modN),\beta+\rho(modN)}=
(\gamma^\rho\Gamma(\gamma^\rho)^{-1})_{\alpha,\beta}=
(\Gamma)_{\alpha,\beta},\label{abel}
\ee
where $\Gamma$ is an arbitrary product of the $\gamma$ matrices.
The action can now be rewritten in a more compact notation as
\be
L^dS[\phi]=\la0,0|\hf\phi(0)[\cp^2\cK(-\cp^2)+m^2]\phi(0)
+{\lambda\over4}\phi^4(0)|0,0\ra
\ee
where the operator $\phi(0)$ is given by
\be
\phi(0)=\int{d^dp\over(2\pi)^d}\phi(p)=\sum_\alpha\int{d^d\tp\over(2\pi)^d}
\phi_\alpha(\tp).
\ee
Elementary rearrangements yield
\be\label{lagrpk}
L^dS[\phi]=tr\Biggl\{\hf\int{d^d\tp\over(2\pi)^d}
\phi\br(-\tp)[\tilde\cK(\tp)+\tilde m^2]\phi\br(\tp)
+{\lambda\over4}\left(\prod\limits_{k=1}^4
\int{d^d\tp_k\over(2\pi)^d}\phi\br(\tp_k)\right)
(2\pi)^d\delta(\sum\limits_kp_k)\Biggr\},
\ee
where $\phi\br=\phi_\alpha\gamma^\alpha$,
and the matrices $\tilde\cK(\tp)$ and $\tilde m^2$ are given by
\bea
\tilde\cK_{\alpha,\beta}(\tp)
&=&\delta_{\alpha,\beta}\cp^2(P^{(N)}(\alpha)+\tp)\cK(-\cp^2(P^{(N)}(\alpha)+\tp)),\nonu
\tilde m^2_{\alpha,\beta}
&=&\delta_{\alpha,\beta}m^2.\label{trband}
\eea
The action \eq{lagrpk} includes $N^{d_{AF}}$ fields whose flavor 
mixing is handled by the matrices $\gamma_\alpha$. In fact, the 
real space expression is 
\be
S[\phi]=\int d^d xtr\biggl\{
\hf\partial_\mu\phi\br(x)\cK\biggl({(2\pi)^2\over\Lambda^2}
\Box\biggr)\partial_\mu\phi\br(x)+V(\phi\br(x))\biggr\}.
\ee
In lattice regularization the lattice sites correspond
to the primitive unit cells of the original vacuum.
Our goal, to trade the momentum non-conservation
on an inhomogeneous vacuum into a multiplicity of excitation bands
over a homogeneous vacuum, is completed.

\section{Elementary excitations}
We determine in this section the elementary excitations which 
are the eigenfunctions of the second
functional derivative of the action, evaluated at the tree level
vacuum. We choose our tree level vacuum in the phase $N$ to be
\be
\phi(x)=\sum_\alpha e^{ix\cdot P^{(N)}(\alpha)}\Phi_\alpha.\label{trvac}
\ee
We need the eigenvectors of
\be
G^{-1}_{\alpha,\beta}(\tp)=
{\delta^2S[\phi]\over\delta\phi_\alpha(-\tp)\delta\phi_\beta(\tp)}
_{\vert\phi=\Phi}
=\left(\tilde\cK(\tp)+\tilde m^2+3\lambda\Phi\br^2\right)_{\alpha,\beta}.
\ee
The propagator can formally be written as
\be
G(\tp)=\sum\limits_{\alpha=1}^{N^{d_{AF}}}\psi_\alpha(\tp)
\lambda_\alpha^{-1}(\tp)\psi^\dagger_\alpha(\tp),
\ee
where
\be
\left(\tilde\cK(\tp)+\tilde m^2+3\lambda\Phi\br^2\right)\psi_\alpha(\tp)
=\lambda_\alpha(\tp)\psi_\alpha(\tp).
\ee
The diagonalization of the quadratic part of the action for a given 
Bloch wave number provides the propagator and the corresponding band 
structure, $\lambda_\alpha(\tp)$.

We start with small values of $c_4$, i.e. we are either in the
ferromagnetic or in the $N=d_{AF}=2$ antiferromagnetic phase, when
$\Phi_\alpha=\delta_{\alpha,1}\phi_{IH}$, or
$\Phi_\alpha=\delta_{\alpha,4}\phi_{IH}$, respectively. 
Since $\gamma_\alpha^2=1$ the inverse propagator is
\be
G^{-1}(p)=m^2_{MF}+\ph^2\cK(-\ph^2),
\ee
where
\be
\ph_\mu=2\sin{p_\mu\over2}
\ee
and
\be
m^2_{MF}=m^2+3\lambda\phi^2_1=\cases{-2m^2&ferromagnetic,\cr
-2m^2-3\cM^2(1,2,d_{AF},c_2,c_4)&antiferromagnetic.}
\label{pmass}
\ee
The propagator can be written as
\be
G^{-1}=m^2_{MF}+\cp^2\left(1-c_2\cp^2+c_4\cp^4\right),\label{propc}
\ee
or
\be
G_{\alpha,\beta}(\tp)=\delta_{\alpha,\beta}G(P^{(2)}(\alpha)+\tp).
\ee
The long wavelength fluctuations give
\be
G^{-1}_{\alpha,\beta}(\tp)=\delta_{\alpha,\beta}\left[
m^2_{MF}(\alpha)+Z(\alpha)p^2+O(p^4)\right]
\ee
where
\be
m^2_{MF}(\alpha)=m^2_{MF}+\cM^2(1,2,n_1(\alpha)+n_2(\alpha),c_2,c_4)
\label{tmass}
\ee
and
\be
Z(\alpha)=\cases{1&$\alpha=1$\cr
-1+8c_2-48c_4&$\alpha=4$}
\ee

The inverse propagator always has a local minimum at $\cp^2=0$.
For certain values of $c_2$ and $c_4$ it may develop another minimum 
at $\cp^2=\cp^2_r>0$ when considered as a function of $\cp^2$.
This minimum is realized kinematically for $\cp^2_r\le8$ only, 
when it is reached on a closed line in the Brillouin zone, c.f. Fig. 
\ref{fourier}, a structure reminiscent of the roton spectrum \cite{land}. 
Thus the van der Waals-type force, represented by the
choice $c_2>0$ leads directly to the appearance of the
additional minima of the dispersion relation interpreted as
rotons \cite{feynm}.

\begin{figure}
\centerline{\psfig{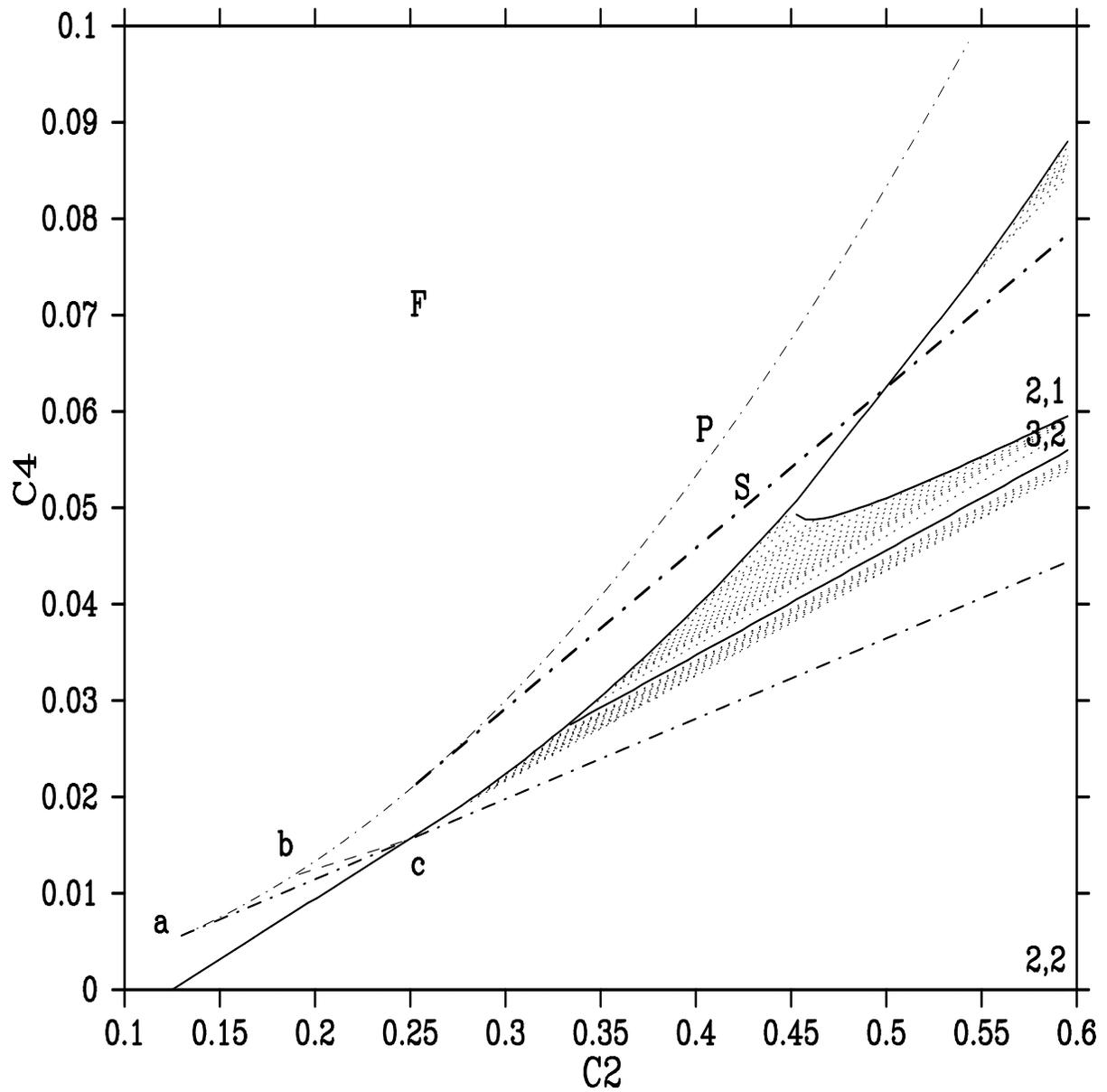}}
\caption{\label{mfpr} 
The zoom into the mean field phase diagram with the regions
characterizing the different behaviour of the propagator. }
\end{figure}

To follow this in detail, c.f. Fig. \ref{mfpr},
we start with the condition for the extremum,
\be
{\partial\over\partial p_\mu}G^{-1}(p)=2\sin p_\mu
\left(1-2c_2\cp^2+3c_4\cp^4\right).\label{extrc}
\ee
Apart of the points $p_\mu=P_\mu^{(2)}(\alpha)$ which are always solutions, the
root of the expression in the parentheses yields another extremum,
\be
\cp^2_r={c_2\over3c_4}\left(1+\sqrt{1-3{c_4\over c_2^2}}\right),\label{secmin}
\ee
so long as 
\be
c_4\le{c_2^2\over3}.\label{qexc}
\ee

The extremum at $\cp^2=\cp^2_r$ is always a minimum.
To select the other minima we need the second derivative matrix,
\be
{\partial^2G^{-1}(p)\over\partial p_\mu\partial p_\nu}
=2\delta_{\mu,\nu}\cos p_\mu\left(1-2c_2\cp^2+3c_4\cp^4\right)
-8\sin p_\mu\sin p_\nu\left(c_2-3c_4\cp^2\right).
\ee
The sub-zones $\alpha=1$ and 4 contain particle like excitations
because the centers of these sub-zones are local minima,
\bea
{\partial^2\over\partial p_\mu\partial p_\nu}G^{-1}(p)_{\vert p=P^{(2)}(1)}
&=&2\delta_{\mu,\nu}\nonu
{\partial^2\over\partial p_\mu\partial p_\nu}G^{-1}(p)_{\vert p=P^{(2)}(4)}
&=&-2\delta_{\mu,\nu}(1-16c_2+192c_4).
\eea
In contrary, the other sub-zones contain saddle point only and
do not support particle like excitations,
\bea
{\partial^2\over\partial p_1\partial p_1}G^{-1}(p)_{\vert p=P^{(2)}(2)}
&=&-{\partial^2\over\partial p_2\partial p_2}G^{-1}(p)_{\vert p=P^{(2)}(2)}\nonu
&=&-{\partial^2\over\partial p_1\partial p_1}G^{-1}(p)_{\vert p=P^{(2)}(3)}\nonu
&=&{\partial^2\over\partial p_2\partial p_2}G^{-1}(p)_{\vert p=P^{(2)}(3)}\nonu
&=&-2(1-8c_2+48c_4).
\eea

When $c_4$ is small enough then $\cp^2_r>8$ and
the inverse propagator possesses two discrete 
minima, at $p=P^{(2)}(1)=(0,0)$ and at $p=P^{(2)}(4)=(\pi,\pi)$. The 
absolute minimum is at
\be
p=\cases{(0,0)&ferromagnetic,\cr(\pi,\pi)&antiferromagnetic.}
\ee
As the value of $c_4$ is gradually increased from zero $\cp^2_r$ reaches the 
value 8 and it is better to follow what happens in the two phases separately.
In the ferromagnetic phase the degenerate local minima are found
along a closed loop in the vicinity of the point $p=(\pi,\pi)$ when \eq{secmin} 
reaches the allowed kinematical regime, $\cp_r^2<8$, for
\be
{c_2\over24}<c_4,~~~{\mathrm and}~~~{c_2\over12}-{1\over192}<c_4.\label{ftrk}
\ee
This line becomes the absolute minimum when the smaller root of the expression
\be
1-c_2\cp^2+c_4\cp^4\label{ringst}
\ee
as the function of $\cp^2$ turns out to be smaller than 8, 
\be
{c_2\over16}<c_4~~~{\mathrm or}~~~c_4<{c_2\over8}-{1\over64}.\label{ftre}
\ee
In fact, \eq{ringst} is negative between the roots and the inverse
propagator goes below its value at $\cp^2=0$ according to \eq{propc}.

It is shown in Fig. \ref{mfpr} that the line of degenerate 
local minima appears within the region bounded by the part ab of the parabola
P, \eq{qexc} and the line segments bc and ac, defined by the first and
the second inequalities of \eq{ftre} and \eq{ftrk},
respectively. The line of degenerate minima turns out to be absolute
minimum below the parabole \eq{qexc} and above the line segment
bc and the ferro- antiferromagnetic phase boundary. The absolute minimum of the
inverse propagator is degenerate and lies on a closed line in either phase
for $c_2>1/4$ when the inequality \eq{ftre} is trivially satisfied.

As the value of $c_4$ is increased in the antiferromagnetic phase
the inverse propagator keeps its minimum at $p=(\pi,\pi)$ so long as the
second inequality of \eq{ftrk} is violated. The absolute
minimum is degenerate and found along a closed loop around $p=(\pi,\pi)$
when $c_4$ is further increased. We may avoid the phase $N=3$ by choosing
large enough $c_2$ and arrive at the $N=2$ nonrelativistic
phase without modifying the propagator at the phase transition. 
The value of the lattice momentum at the minimum, \eq{secmin}, is a 
monotonically decreasing function of $c_4$ and reaches 4 for
\be
c_4={c_2\over6}-{1\over48},\label{square}
\ee
the line S of Fig. \ref{mfpr}. 
At this point the degenerate minima of the inverse propagator
form a square and for larger values of $c_4$ it is deformed
into a closed loop around $p=0$.

The further increase of $c_4$ brings us to the higher $N$ phase boundaries
and the situation becomes more involved.

We turn now to the question of the critical points. According to 
\eq{pmass} and \eq{tmass} the second order phase transition
is reached when both $m^2\to0$ and $\cM^2\to0$ for $m^2<0$.
Let us take 
\be
\cM^2=-\mu^2_{\cM}a^2,~~~~~m^2=-\mu_m^2a^2.\label{trfint}
\ee
The first equation and \eq{propg} assert that the criticality is
reached either for
\be
c_4\approx{c_2\over4d_{AF}\sin^2\left({K\over2}\right)}
-{1\over16d^2_{AF}\sin^4\left({K\over2}\right)},
\ee
or as $N/M\to\infty$. Both cases require the vicinity of the 
ferro- antiferromagnetic transition line, the value of 
$c_2$ and $c_4$ is finite in the first case and diverging in
the second. Since $C_4(N)\ge C_2^2(N)$ \eq{actaf} shows that
$\mu^2_{\cM}>0$ is needed to reach this phase transition. Thus
the ferro-antiferromagnetic transition line is critical
in the mean-field approximation.

We close this Section with a remark about the Goldstone modes.
The inhomogeneous vacuum breaks the external symmetries and we find
Goldstone modes for the models in the continuum. The lattice regulator
reduces the space-time symmetries into a discrete group and there
is no reason to expect massless phonons in the antiferromagnetic
phases. But the strength of the breaking of the continuous part of the
space-time symmetries is $O(M/N)$. Thus the continuous
kinematical symmetries are restored as $N/M\to\infty$ 
and the Goldstone theorem is regained asymptotically, i.e. 
$\lambda(p_{min})=O(M/N)$. An important consequence
of this gradual restoration of the Goldstone theorem is the
absence of long range modes and the possibility of supporting 
the periodic vacuum in two dimensions when $N$ is finite. This
does not mean a long range order because the non-trivial part of the
vacuum is squeezed within the distance $Na/M\to0$.

\section{One-loop renormalization}
It is shown in this Section that the one-loop effective potential of our model 
can be made finite by the introduction of an appropriate mass counterterm and
the ferro-antiferromagnetic transition remains a critical line on the
one-loop level.

It is straightforward to derive the Feynman rules for \eq{lagrpk}
and to compute the radiative corrections. The one-loop effective 
potential, the generating function for the 1PI vertices,
\be
V_{eff}(\Phi)=\sum_{n=0}^\infty{1\over n!}
\sum_{\alpha_1,\cdots,\alpha_n}\Phi_{\alpha_1}\cdots
\Phi_{\alpha_n}\Gamma^{(n)}(P^{(N)}(\alpha_1),\cdots,P^{(N)}(\alpha_n)),
\ee
can be written in the one-loop approximation as
\be
V_{eff}(\Phi)=V_{tree}(\Phi)+\hf\int{d^d\tp\over(2\pi)^d}
\tr\ln[\tilde\cK(\tp)+\tilde m^2+3\lambda\Phi\br^2],\label{heffpot}
\ee
where
\be
V_{tree}(\Phi)=\tr\left(\hf\tilde\cK(0)\Phi\br^2
+\hf m^2\Phi\br^2+{\lambda\over4}\Phi\br^4\right).
\ee
We split the mass term into a renormalized and a counterterm,
\be
m^2=m^2_R+\delta m^2,
\ee
and make the replacement $m^2\to m_R^2$ in the loop integral.

The $\Phi$-dependence of the one-loop integrals can significantly 
be simplified by using the relation \eq{abel} with $\Gamma=\Phi\br^2$,
\be
(\Phi\br^2)_{\alpha,\alpha}={1\over N^{d_{AF}}}\tr\Phi\br^2,
\ee
where no summation is made for the index $\alpha$.
Since the matrix $\tilde K(\tp)+\tilde m^2_R$ is diagonal,
\be
V_{eff}(\Phi)=V_{tree}(\Phi)+\hf\int{d^d\tp\over(2\pi)^d}
\tr\ln\left[\tilde\cK(\tp)+\tilde m^2_R
+{3\lambda\over N^{d_{AF}}}\tr(\Phi\br^2)\right].\label{eeffpot}
\ee

The ultraviolet divergences are identified in two steps. First we expand 
in the field dependence and write the loop contribution as
\be
\sum_n{1\over2n}\int{d^d\tp\over(2\pi)^d}
\tr\left({3\lambda \tr\Phi\br^2\over\tilde\cK(\tp)+\tilde m^2_R}\right)^n,
\ee
to recover the usual one-loop resummation in the
effective potential. In the second step we expand the integrals
in the lattice spacing $a$,
\bea
a^{2n-d}\int\limits_{|p|<\pi/N}{d^d\tp\over(2\pi)^d}
\tr\left(\tilde K(\tp)+\tilde m^2_R\right)^{-n}
&=&\int\limits_{|p|<\pi/Na}{d^d\tp\over(2\pi)^d}
\biggl[\sum_\alpha{1\over a^2}\cp^2\left(a^2(P^{(N)}(\alpha)+\tp)\right)\nonu
&&\times K\left(-{1\over a^2}\cp^2\left(a^2(P^{(N)}(\alpha)+\tp)\right)\right)
+{\tilde m^2_R\over a^2}\biggr]^{-n}.
\eea
We have at most a logarithmic divergence in two dimensions ($n=1$)
and the dimensionless inverse propagator, the 
integrand of the left hand side, has a finite, $1/a$ independent minimum.
Let us write the smallest eigenvalue of the inverse propagator
around its minimum as
\be
\lambda_{\alpha_{min}}(\tp_{min}+\tp)=m^2_4+Z_0\tp^2+O(a^2\tp^4),
\ee
with $m^2_0=a^2m_{min}^2$,
what allows us to identify the divergent part of the loop integral for $d=2$,
\be
\hf\int\limits_{|p|<\pi/Na}{d^2\tp\over(2\pi)^2}
{1\over m^2_{min}+Z_0\tp^2}.
\ee
Thus the one loop ultraviolet divergence can be removed by setting
\be
\delta m^2=-\int\limits_{|p|<\pi/Na}{d^2\tp\over(2\pi)^2}
{1\over m^2_{min}+Z_0\tp^2}.
\ee
When $m^2_{min}=O(a^{-\epsilon})$, $\epsilon>0$, then no divergency arises. 
If there are several finite minima of the dimensionless inverse propagator
then we sum over them in the counterterm.

The continuum limit, \eq{trfint} can be achieved along 
the ferro-antiferromagnetic transition line of the plane 
$(c_2,c_4)$. We have one or two  particles by approaching this line from 
the ferro- or the antiferromagnetic phase below the point 
$c=(1/4,1/64)$, respectively. Above the point c the rotons appear in both phases and 
replace one of the particles of the antiferromagnetic vacuum. 
One can find specially interesting continuum limits in the
vicinity of the point c. (i) Approach from below the line ab in the ferromagnetic phase,
which gives a single particle and monotonically increasing inverse
propagator with the momentum. (ii) The approach above the line ab but
below bc yields one particle and rotons. The roton momentum diverges with 
the cutoff, a reminescent of the new excitation bands in the antiferromagnetic 
phase. The value of the inverse propagator at the degenerate
minima, the roton mass square
is finite but larger than for the absolute minimum, at $p=0$. 
Thus the rotons are heavier than the particle.
(iii) Approaching above the line bc yields a model
with a particle and rotons where the rotons are lighter than the particle.
(iv) The approach from the antiferromagnetic phase produces a model
with two particles.

\section{Non-Perturbative Vacuum}\label{montec}
\begin{figure}
\centerline{\psfig{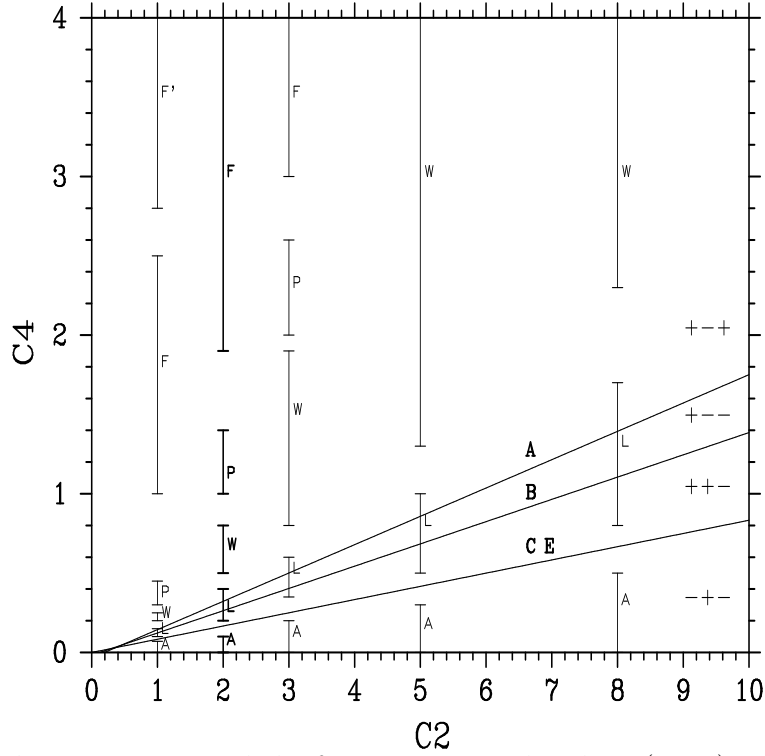}}
\caption{\label{frustf} The phase structure and the frustrations
on the plane $(c_2,c_4)$.
The lines where the coefficients $A$, $B$, $C$ and $E$ of the 
lattice action change sign in the $(c_2,c_4)$ plane. The letters along the
vertical lines are to indicate the qualitative space-time 
structure of the vacuum seen in the simulation: 
A=ordered $N=2$ antiferromagnetic; L=labyrinths; W=plane waves; 
P=weakly antiferromagnetic (onset of the crossover on the finite lattice),
F=ferromagnetic and F'=weakly ferromagnetic.}
\end{figure}

In order to assess the importance of the fluctuations, we performed 
a Monte Carlo simulation of the model \eq{res}. The 
resulting phase structure is summarized on Fig. \ref{frustf}.
The coupling constants $\lambda=0.05$, $m^2=-0.1$ 
were chosen and the simulation on the line $c_2=2$ was done at
$200\times200$ lattice size. The $c_4$ dependence of the results
was monitored with a particular care by scanning
the region $0\le c_4\le1$ with step size $\Delta c_4=0.01$. 
The other points were obtained on $40\times40$ lattice. 
The performed tests have showed no appreciable finite size 
dependence in the qualitative and quantitative features of the 
phase diagram. We used the metropolis update algorithm,
carefully checking the statistics to make sure that no
statistical error would change our conclusions
concerning the phase diagram. In order to test ergodicity, local 
cluster algorithms were tried as well as different initial 
conditions. 

The letters along the vertical lines of Fig. \ref{frustf} indicate 
the qualitative space-time structure of the vacuum seen in the simulation: 
A=ordered ($N=d_{AF}=2$) antiferromagnetic; L=labyrinths; W=plane waves; 
P=weakly antiferromagnetic (onset of the crossover on the finite lattice),
F=ferromagnetic and F'=weakly ferromagnetic. To understand the phase
structure we recall that the parameters of the bare action characterize 
the dynamics at the cutoff. Consequently one may extract useful 
informations about the short range order of the vacuum by considering 
the coefficients appearing in the quadratic part of \eq{res}.
The sign of the coefficients determines the ferro- or antiferromangetic
nature of the couplings. We introduce the subscript $(n,m)$ which shows 
the separation of the two field variables multiplied by the given parameter,
$A_{0,0}=2+m^2/2-10c_2+56c_4$, $B_{1,0}=-1+8c_2-57c_4$, 
$C_{2,0}=-c_2+12c_4$, $E_{1,1}=-2c_2+24c_4$, 
$D_{3,0}=-c_4$ and $F_{1,2}=-3c_4$.
The lines where $A$, $B$, $C$ and $E$ change sign,
\bea
c_4&=&{10\over56}c_2-{m^2\over112}-{2\over56}~~~(A),\nonu
c_4&=&{8\over57}c_2-{1\over57}~~~(B),\nonu
c_4&=&{1\over12}c_2~~~(C,E)
\eea
are shown in Fig. \ref{frustf} by solid lines. These lines intersect
and the short range order varies in a complicated manner for $c_2<0.45$.
But once this value is reached, the sequence of the change of the signs as $c_4$ 
increases is always the same. The importance of these signs is that
they create frustrations whenever $c_4>0$. 
There are no frustration when $c_4=0$ because 
$D_{3,0}=F_{2,1}=0$ and all others favor the $N=2,~d_{AF}=2$
antiferomagnetic or the ferromagnetic vacuum for $c_2>1/8$
or $c_2<1/8$, respectively. For $c_4>0$ but below the line $C,E$, in the 
antiferromagnetic $N=2,~d_{AF}=2$ phase $B_{1,0}>0$, $C_{2,0},~E_{1,1}<0$ 
favor this kind of vaccum. But the other signs, $D_{3,0},~F_{2,1}<0$
introduce frustrations whose density increases with $c_4$. 
The further increase of $c_4$ flips the sign
of $B$ and ultimately $A$, destabilizes the lock-in mechanism at
$N=2$ and opens the way for the rapid variations
of the devil's staircase. 
In between the lines $C,E$ and $B$, $C_{2,0},~E_{1,1}>0$ and 
the frustration density is increased because only $B_{1,0}$
favors this vacuum. The result is a strong increase of the
fluctuations. The mean-field approximation is obviously unreliable 
in this regime and the simulation produces labyrinth-like vacuum, see below. 
For large $c_4$, in the ferromagnetic phase $B_{1,0},~D_{3,0},~F_{1,2}<0$ 
act in favor of the homogeneous vacuum but $C_{2,0},~E_{1,1}>0$ 
generate frustrations which might explain the weakening of the
ferromagnetic condensate for large $c_4$. For the intermediate
values of $c_4$ the competition between the different terms
is more involved and the compromise between the different
competing terms is reached over a longer length scale according 
to the mean-field solution. The qualitative conclusion is the
separation of the stable low $c_4$ $N=2,~d_{AF}=2$ vacuum from the
high $c_4$ ferromagnetic and an intermediate $c_4$ strongly 
frustrated antiferromagnetic vacuua.

The frustrations are the lattice defects of the antiferromagnetic vacuum.
In order to understand their production mechanism in the $N=2$
antiferromagnetic phase we used two different initial 
conditions, an ordered and a disordered one,
\be
\phi^{ord}_{init}(x)=\phi_{IH}(-1)^{x^1+x^2},\label{afdom}
\ee
and 
\be
\phi^{dis}_{init}(x)=0,
\ee
respectively. The initial condition $\phi^{dis}_{init(x)}$ which actually
looks ordered becomes disordered after few Monte-Carlo sweeps. This happens 
because it represents an unstable equilibrium position and the local field
variable "rolls down" to one of the potential minima $\phi(x)=\pm\phi_{IH}$. 
This leads to the formation of $N=2$ antiferromagnetic domain structure.
These domains are separated by walls of links where
the field variable has the same sign. The region on the plane $(c_2,c_4)$
where the antiferromagnetic domains consisting of the patches 
$\phi^{ord}_{init}(x)$ and $-\phi^{ord}_{init}(x)$ develop long
and winding boundaries is called labyrinth-like and is denoted by
L in Fig. \ref{frustf}.

For large values of $c_2$ the amplitude of the
modulated vacuum is large so the frustrations move very slowly in the 
simulation time. The antiferromagnetic domain walls turned out to be
very slow variables, as well. The domain walls were always generated 
below the line $C,~E$ of Fig. \ref{frustf} when the disordered initial 
configuration was used. The
vacuum obtained by the runs with ordered initial configuration 
did not support the domains. The question is whether the domain walls
are real degrees of freedom or reflect the insufficient
convergence of the simulation method.
We have developed cluster algorithm and found that for small $c_2$,
($c_2<1.0$), the walls have dynamics, reach an equilibrium and may disappear.
For larger $c_2$ the domain wall motion slows down despite of the
cluster algorithm. The 
thermalization was safely reached within each domain.
It remains an intriguing question if the domain walls
thermalize in this regime with extremely long relaxation time, 
i.e we are in a glassy regime \cite{spgl} or the ergodicity is 
definitely lost and the vacuum consists of a stable,
disordered network of domain walls. In this regime the frustrations
act as scattering centers without feedback from the fast dynamics of the
elementary excitations, a dynamical situation reminiscent of the quenched
disorder in solids. For $c_2<1$ the mean field value, $\phi_{IH}$,
is small enough to make the domain wall fluctuations more likely
and the cluster update averages over the different rearrangements
of the domains. 

\begin{figure}
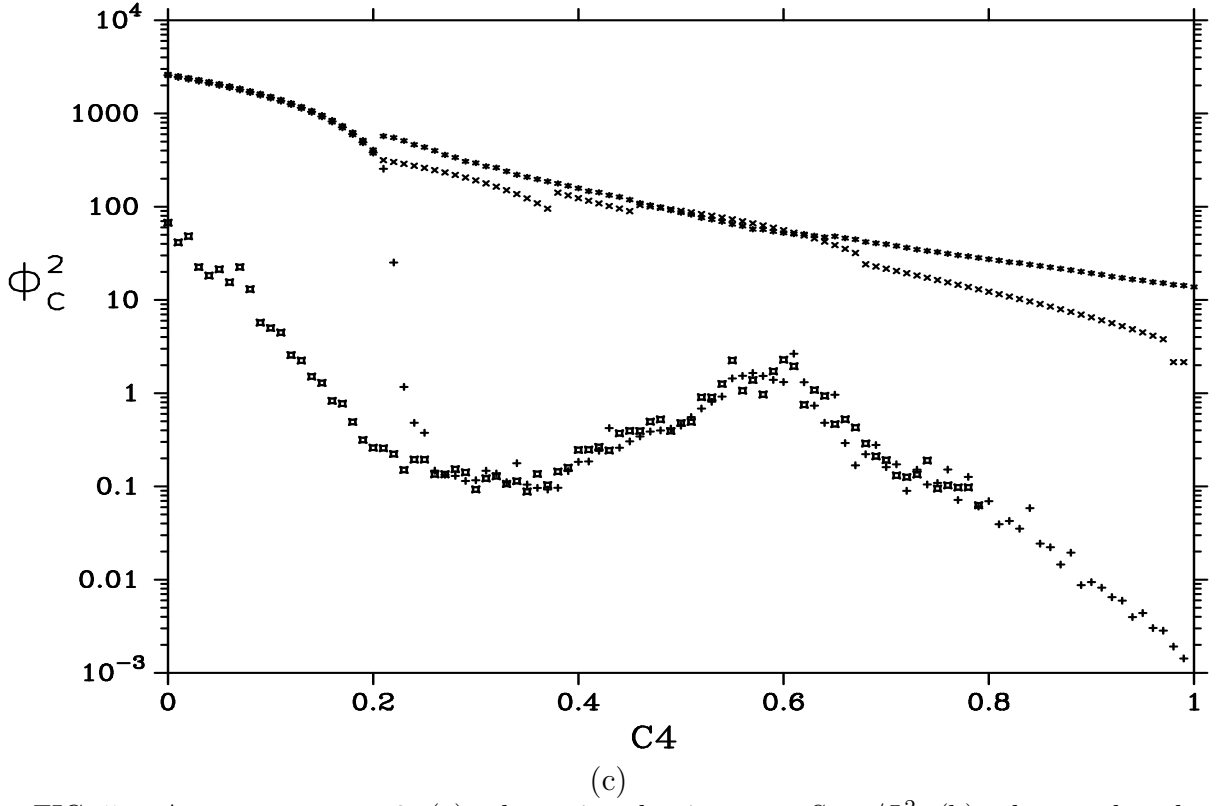

\centerline{\psfig{file=Fig4a.ps,height=10cm,width=16cm,angle=270}}
          \centerline{(a)}
\centerline{\psfig{file=Fig4b.ps,height=10cm,width=16cm,angle=270}}
          \centerline{(b)}
\centerline{\psfig{file=Fig4c.ps,height=10cm,width=16cm,angle=270}}
          \centerline{(c)}
\caption{\label{nonpv}
Averages at $c_2=2$, (a): the action density, $-<S>/L^2$, 
(b):  the wavelength of the
modulation of the vacuum, measured by $-<\phi\Box\phi>/<\phi^2>$, 
(c): the amplitude of the modulation.}
\end{figure}

The different regions shown in Fig. \ref{frustf} were studied in more
detail at $c_2=2$. 
The negative action density obtained by starting at $c_4=0$ and sweeping
the interval $0\le c_4\le1$ is plotted in Fig. \ref{nonpv}(a). The 
results obtained by the ordered and disordered initial configuration 
are indicated by + and squares, respectively. The ordered
vacuum has lower action density up to $c_4\approx0.2$. 
The mean-field action density is shown by a solid line. It is
instructive to follow the second lowest mean-field solution,
indicated by the circles. The splitting between the lowest and
the second lowest mean field action level reflects the stability
of the vacuum against the change of the long range order.
The agreement between the Monte Carlo and the mean-field results is remarkable
for small $c_4$ in the relativistic phase. Right at the
relativistic-non relativistic phase transition, the ordered 
configuration is a rather poor approximation
and the disordered vacuum adjusts itself easier to the value $d_{AF}=1$.
For $c_4>0.3$ the mean field badly over estimates the true 
vacuum action density, indicating the presence of strong fluctuations.
Notice that the phases with $N=2$ and 4 are more stable against
the modification of the long range, in agreement with their
stronger "lock-in" in Fig. \ref{devil}. 

Another insight into the vacuum can be gained from the inspection of
the period length of the vacuum, $\lambda_{vac}$, measured by
\be
-{<\phi\Box\phi>\over<\phi^2>}\approx 
4d_{AF}\sin^2\left({\pi\over\lambda_{vac}}\right).\label{wvl}
\ee
The numerical and the analytical results depicted in Fig. \ref{nonpv}(b)
show that the first order transitions of the mean field approximation
between the relativistic and the non-relativistic $N=2$ antiferromagnetic
phase is smoothened out in the simulations when the disordered initial 
condition is used. The ordered initial condition follows the mean field
curve within the phase $N=d_{AF}=2$. The slightly higher action
of the disordered initial configuration runs indicates that the true
vacuum is close to being ordered and the relativistic-non relativistic
phase transition is of strongly first order. One is tempted to conclude that
the fluctuations smoothen out the commensurate-incommensurate transitions
but better statistics is needed to settle this question in a satisfactory
manner for the whole phase diagram. The difference between the numerical 
results and the mean-field solution is the largest in the phase
$N=2,~d_{AF}=1$. The mean-field 
approximation slightly underestimates the period length of 
the vacuum in the vicinity of the ferromagnetic transition. This is
consistent with one of the remarks made about Fig. \ref{nonpv}(a), namely
that the fluctuations in this regime are stronger than expected by 
the mean-field approximation. In fact, the stronger fluctuations lower the
critical value of $c_4$ so the period length of the vacuum diverges faster in
function of $c_4$ than in the mean-field expression.

The strength of the modulation of the vacuum is displayed in Fig. \ref{nonpv}(c).
It is simplest to express it in terms of the Fourier transformed field
\be
\tilde\phi(p)={1\over L^2}\sum_x\phi(x)e^{-ip\cdot x},
\ee
obtained on an $L\times L$ lattice.
We may split the expectation value $<|\tilde\phi(p)|^2>$ into the sum of 
the condensate and the fluctuations, 
\be
<|\tilde\phi(p)|^2>=<|\tilde\phi(p)|^2>_c+<|\tilde\phi(p)|^2>_{fl},
\label{sepe}
\ee
\be
<|\tilde\phi(p)|^2>_c=|<\tilde\phi(p)>|^2,~~~
<|\tilde\phi(p)|^2>_{fl}=G_c(p).\label{sepk}
\ee
were $G_c(p)$ stands for the connected propagator, given by 
\eq{propc} in the leading order of the perturbation expansion. 
A simple approximation for the strength of the modulation of the vacuum is
\be
\phi_c^2=max_p\la|\tilde\phi(p)|^2\ra_c\approx max_p\la|\tilde\phi(p)|^2\ra
\approx\la max_x\phi(x)\ra^2\approx\la min_x\phi(x)\ra^2.\label{condmax}
\ee
$max_p\la|\tilde\phi(p)|^2\ra$ is displayed by plus and square for the
ordered and disordered initial configuration, respectively. The star
shows the value of the last two expressions in \eq{condmax}. We find
that the different estimates for the
strength of the condensate agree in the relativistic phase
except the simulation results corresponding to the disordered start. 
The difference between the mean-field solution indicated with x
and the numerical results can
be considered as a measure of the strength of the fluctuations.
Note the local maximum and minimum in the fluctuations at $c_4\approx0.35$
and $c_4\approx 0.6$, respectively. The former is in agreement with
the remark made for Fig. \ref{nonpv}(b).  
The average of the extrema of the field in the real space
is higher than the mean field value and shows no structure as the function of
$c_4$ indicating that the fluctuations tend to lower the local values of 
the field variable $|\phi(x)|$. As noted before, the fluctuations increase as the 
ferromagnetic phase is approached, a result consistent with the
second order nature of the ferromagnetic phase transition. 

\begin{figure}
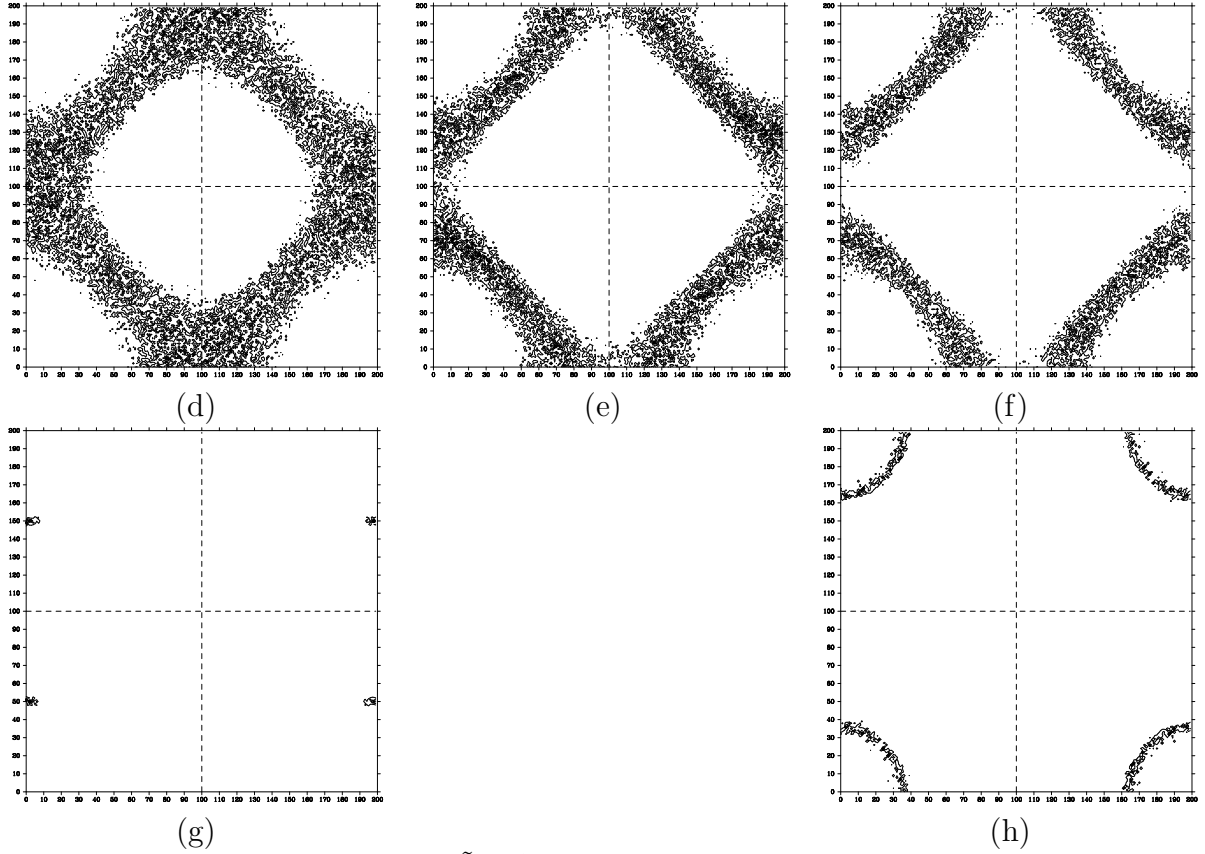

\begin{minipage}{5cm}
\centerline{\psfig{file=Fig0.06o.ps,height=5cm,width=5cm,angle=270}}
          \centerline{(a)}
\end{minipage}
\hfill
\begin{minipage}{5cm}
\centerline{\psfig{file=Fig0.23o.ps,height=5cm,width=5cm,angle=270}}
          \centerline{(b)}
\end{minipage}
\hfill
\begin{minipage}{5cm}
\centerline{\psfig{file=fig0.23.ps,height=5cm,width=5cm,angle=270}}
          \centerline{(c)}
\end{minipage}

\begin{minipage}{5cm}
\centerline{\psfig{file=fig0.31.ps,height=5cm,width=5cm,angle=270}}
          \centerline{(d)}
\end{minipage}
\hfill
\begin{minipage}{5cm}
\centerline{\psfig{file=fig0.38.ps,height=5cm,width=5cm,angle=270}}
        \centerline{(e)}
\end{minipage}
\hfill
\begin{minipage}{5cm}
\centerline{\psfig{file=fig0.39.ps,height=5cm,width=5cm,angle=270}}
        \centerline{(f)}
\end{minipage}

\begin{minipage}{5cm}
\centerline{\psfig{file=fig0.57.ps,height=5cm,width=5cm,angle=270}}
        \centerline{(g)}
\end{minipage}
\hfill
\begin{minipage}{5cm}
\centerline{\psfig{file=fig0.90.ps,height=5cm,width=5cm,angle=270}}
          \centerline{(h)}
\end{minipage}
\caption{\label{fourier}
The Fourier transform $<|\tilde\phi(p)|^2>$ in the plane 
$(p_1,p_2)$ for $c_2=2$. 
Disordered initial conditions: (a): $c_4=0.06$; (b): $c_4=0.23$.
Ordered initial conditions: (c): $c_4=0.23$; (d): $c_4=0.31$;
(e): $c_4=0.38$; (f): $c_4=0.39$; (g) $c_4=0.57$; (h): $c_4=0.9$.}
\end{figure}

The numerical results for the complete propagator, \eq{sepe}
what displays the structure of the vacuum and the elementary
excitations in the same time are presented in Fig.
\ref{fourier}. We divided the interval $[0,max_p\la|\phi(p)|^2\ra]$
into five equal segments and their contourplots are shown in the figures. 
The strength of the contour line increases with the amplitude, so 
the blacker regions of the plots indicate the location of the maxima.

\underbar{$0<c_4<0.16$:} 
When $c_4$ is small we are in the relativistic $N=2$ phase and
$<|\phi(p)|^2>$ depends strongly on the initial condition
of the Monte Carlo simulation. The ordered initial configuration 
\eq{afdom} yields a single peak for $<|\tilde\phi(p)|^2>$ 
at $p=P^{(2)}(4)=(\pi,\pi)$ suggesting little disorder.
In the case of the disordered initial configuration, $\phi_{init}^{dis}$,
one finds a domain structure and
the Fourier transform $|<\tilde\phi(p)>|^2$ shown in Fig. \ref{fourier}(a)
at $c_4=0$ is depleted at $p=P^{(2)}(4)$,
$<|\phi(P^{(2)}(4))|^2>\approx0$, and develops a ring of maxima around 
this momentum. It is easy to understand the minimum at the center.
In fact, assuming that the vacuum consists of the domains of $\phi^{ord}_{init}(x)$
and $-\phi^{ord}_{init}(x)$ in the fractions $c$ 
and $1-c$ of the volume, respectively one finds
$\phi_{cond}^2=\phi_{IH}^2(1-2c)^2/4$. 
The domain pattern develops after few sweeps and the domain walls
turn out to be rather slow variables. 

\underbar{$0.16<c_4<0.22$:} 
The excitations become more involved in this regime. The coefficients
$C$ and $E$ of the lattice action are positive for $c_4>1/6$
making the frustration density higher. Furthermore the propagator 
develops a circle of degenerate maxima around the point $p=P^{(2)}(4)$ 
for $c_4>31/192\approx0.16$. The result is a cusp in the condensate 
as the function of $c_4$ cf. Fig. \ref{nonpv}(c), and the softening 
of the modes giving an increase of $<|\tilde\phi(p)|^2>$.

\underbar{$0.22<c_4<0.3$:} 
The ordered initial configuration simulation recovers the right vacuum at 
$c_4\approx0.23$ in a discontinuous
manner, c. f. Fig. \ref{fourier}(b) and (c). 
For $c_4>0.23$ the simulations corresponding to the two different
initial conditions yield the same result. 
The hysteresis in the $c_4$ dependence, i.e. the later appearance of the
ring for the ordered initial configuration case compared with 
the unstable starts suggests that the transition 
$d_{AF}:1\longleftrightarrow2$ is of first order. 
The roton minimum in the
dispersion relation tends to break the straight lines where the
frustrations are found and to distribute them in a more spherically
symmetrical manner, resulting in more disorder and creating a
labyrinth structure instead of the ordered nonrelativistic 
antiferromagnetic vacuum. 

\underbar{$0.3<c_4<0.38$:} The minimum of \eq{propc} 
is the longest, being a square, for $c_4=0.31$ according to \eq{square}.
A typical example shown in Fig. \ref{fourier}(d) witnesses that the 
fluctuations are the strongest in this regime, when the volume of the 
phase space occupied by the soft modes is the largest.
As the length of the roton minima starts to shrink for $c_4>0.31$
so does the strength of the fluctuations as seen in Fig. \ref{nonpv}(c).

\underbar{$0.38<c_4<0.6$:} 
The period length of the vacuum growths beyond 2 at $c_4\approx0.38$ 
and $<|\tilde\phi(p)|^2>$ becomes strongly suppressed at 
$p=(0,\pi)$ and $p=(\pi,0)$ according to Fig. \ref{fourier}(e)-(f).
The rapid variation in $c_4$ supports the discontinuous nature of the
transition  $N=2\to3$. The further increase of $c_4$ 
makes the restoration of the rotational symmetry more 
difficult. This is because the symmetry restoration is achieved by
forming domains where the plane wave modulation of the vacuum has
different orientation. The longer period length of the vacuum 
makes more energy consuming to break the straight plane wave
by changing its direction, i.e. the domain wall energy density 
increases.
This explains the breaking up of the ring into smaller segments
and its ultimate reduction to few discrete peaks, as
demonstrated in Fig. \ref{fourier}(g). 

\underbar{$c_4\approx0.6$:} 
The fluctuations reach a minimum in the middle of the "lock-in" 
interval $N/M=4$ seen in Fig. \ref{devil}.
The point where $N/M$ is approximately the power of the highest order
terms in the field variable of the lagrangian the above mentioned
decrease of the fluctuations comes to a halt and is turned
into an opposite trend due to the approach of the ferromagnetic
phase transition. In fact, as the length of the roton minima shrinks and becomes 
less important the increase of $c_4$ ultimately leads to the disappearance
of the condensate which is triggered by the increase of the fluctuations.
This is due to a tree-level effect, $c_4$ makes the modulation of
the vacuum more energetic so the amplitude of the modulation
decreases with the increase of $c_4$. As the amplitude
decreases it becomes easier to break a plan wave into domains what amounts
to the attempt for the restoration of the rotational symmetry. The maximum
in $<\tilde\phi(p)^2>$ spreads from a well localized point over the whole circle 
of the rotons as $c_4$ is increased beyond $0.6$.

\underbar{$0.6<c_4<0.98$:} 
As $c_4$ increases and $p_{min}$ approaches zero the dominant fluctuations 
are grouped on a circle around $p=0$ with increasing strength as shown in
Fig. \ref{fourier}(h). The 
condensate weakens and increasing period length in lattice 
spacing units is in agreement with the one-loop renormalizability
established in the previous section.

\underbar{$c_4\approx0.98$:} 
The precursor of the transition to the ferromagnetic phase
is the appearance of a peak in $<|\tilde\phi(p)|^2>$ at $p=0$
for $c_4\approx0.98$. The further increase of $c_4$ brings us into
the ferromagnetic phase with roton excitations.

We close this section with a remark concerning the Lifshitz point.
It is a tricritical point at $g_2=c_1=0$ in the model \eq{lipol} where
the wave vector of the periodic vacuum tends to zero as $c_1\to0^-$.
This is to be compared with the "Lifshitz line" of our model,
the curve separating the ferromagnetic 
and the antiferromagnetic phases in Fig. \ref{mfphstr}. For a given
$c_4\ge0$ this line gives a critical point which is reached by tuning
$c_2$ and $m^2$. Since the coefficient of the lowest order term of the gradient
in the action is kept constant ($c_1=1$) the fine tuning of the higher
order coefficients generates discontinuity for the wave vector of
the vacuum at the critical point\footnote{The analogous situation
at the usual tricritical point for the model \eq{lipol} with $c_2=0$,
$g_6>0$ is the first order phase transition in the function of $g_4$.
We do not need $c_4>0$ in our case because the anharmonic term of the
lagrangian stabilizes the vacuum for $c_2<0$, $c_4=0$.}. The peculiarity
of the extension of the Lifshitz point to a line is that the wave
vector of the vacuum is a discontinuous function of the coupling
constants either when one crosses the line or when one moves
along it in the antiferromagnetic side. One point of the Lifshitz
line shows a further interesting feature, the even and the odd 
sublattices decouple at $c_2=1/4d$, $c_4=0$. This can be understood
by checking the invariance of the propagator under the replacement
$\cp^2\to 4d-\cp^2$, and allows us to construct the
continuum limit of chiral bosons \cite{ehj}.

\section{Wick rotation to real time}
Our model \eq{lagr} was obtained by the elimination of
some particles and the truncation of the resulting effective action.
The usual strategy to avoid the instabilities in real time
mentioned above is based on the introduction of new, auxiliary fields which render
the kinetic term quadratic. The most natural attempt is suggested
by the formal similarity between the Pauli-Villars regulated models
and theories with higher order derivatives \cite{pauh}, \cite{hiderk}. 
In fact, by assuming that the 
inverse propagator \eq{inprop} as a function of $p^2$ has
simple roots only, we can write the propagator as
\be
G(p^2)={1\over(p^2+M^2_1)(p^2+M^2_2)(p^2+M^2_3)}
={Z_1\over p^2+M_1^2}+{Z_2\over p^2+M_2^2}
+{Z_3\over p^2+M_3^2},
\label{pvprop}
\ee
where
\bea
Z_1^{-1}&=&(M_2^2-M_1^2)(M_3^2-M_1^2),\nonu
Z_2^{-1}&=&(M_1^2-M_2^2)(M_3^2-M_2^2),\\
Z_3^{-1}&=&(M_1^2-M_3^2)(M_2^2-M_3^2).\nonu
\eea
Since 
\be
Z_1^{-1}Z_2^{-1}Z_3^{-1}=-(M_2^2-M_1^2)^2(M_3^2-M_1^2)^2(M_3^2-M_2^2)^2,
\ee
one of the coefficients $Z_j$ is negative when the roots $M_j^2$ are real,
i.e. in the case of the antiferromagnetic vacuum. We have two complex roots,
say $M_2^2=M_3^{2*}$ in the ferro- or the paramagnetic phase. This gives
$Z_2^{-1*}=Z_3^{-1}$ and
\be
Z_2^{-1}+Z_2^{-1*}=(M_3^2-M_2^2)^2<0,
\ee
and makes $ReZ_2=ReZ_3<0$. Thus the real part of at least 
one of the contributions in \eq{pvprop} is always negative.

Such a propagator can formally be obtained as the functional derivative of
a Gaussian generator functional,
\be
Z_0[j(x)]=\int D[\phi_j]e^{-\hf\int d^dxd^dy\phi_j(x)G_{j,k}(x,y)\phi_k(y)
+\int d^dxj(x)\sum\limits_{k=1}^3\phi_k(x)},
\label{pavi}
\ee
where
\be
G_{j,k}(x,y)=-\delta(x-y)\delta_{j,k}
Z^{-1}_j(\Box_y-M^2_j).
\ee
Note that the convergence of the
functional integration requires that the field $\phi_j(x)$ with 
$ReZ_j<0$ be purely imaginary, an indication of the presence
of the negative norm states in the canonical quantization procedure.
The partition function of the model \eq{lagr} can be written as
\be
\int D[\phi]e^{-S[\phi]}=e^{-\int d^dxV
\bigl({\delta\over\delta j(x)}\bigr)}Z_0[j]
\ee
in the framework of the perturbation expansion. The Wick rotation of this
path integral is straightforward and leads to a model with three particles
and negative norm Hilbert space. The positive and the negative
norm particles are mixed by the vertices of the interaction $V(\phi)$.
In order to have a physically acceptable
model we have to ensure that the time evolution remains unitary 
when restricted to the subspace of positive norm physical states
and the Hamiltonian is bounded from below. 

The unitarity of the
S-matrix can be established for energies below the threshold
of the negative norm particle production, the key element for 
the applicability of the Pauli-Villars regularization. The stability
of the vacuum is reached in this regularization
scheme by first performing the 
renormalization in the Euclidean space and making the Wick
rotation back to real time after that. The
negative norm regulator particles are suppressed during the
removal of the cutoff thereby the Wick rotated
renormalized theory possesses a stable ground state. The
non-commutativity of the renormalization and the Wick rotation
can simply be understood by noting that the formally Lorentz
invariant Pauli-Villars regularization scheme violates 
a simple rule: The non-compact
nature of the Lorentz group excludes the Lorentz invariant
regulator schemes. 

What we have shown so far is that our 
model yields an acceptable theory for real time
only after the perturbative renormalization in Euclidean space-time.
One wonders if canonical quantization which leads to a non-perturbative
formalism can carry us further in establishing an acceptable,
non-perturbative theory. 
In our case it amounts to the introduction of the generalized
coordinates $\phi_j(x)=\partial^j_0\phi(x)$, $j=0,1,2$, and momenta
\bea
\Pi_0&=&{\delta L\over\delta\partial_0\phi}
-\partial_0{\delta L\over\delta\partial_0^2\phi}
+\partial^2_0{\delta L\over\delta\partial_0^3\phi}\nonu
&=&\left[1-c_2{(2\pi)^2\over\Lambda^2}\Box
+c_4\left({(2\pi)^2\over\Lambda^2}\right)^2
(\Box^2-\Box\nabla^2)\right]\phi_1,\nonu
\Pi_1&=&{\delta L\over\delta\partial_0^2\phi}
-\partial_0{\delta L\over\delta\partial_0^3\phi}\nonu
&=&\left[c_2{(2\pi)^2\over\Lambda^2}\Box
-c_4\left({(2\pi)^2\over\Lambda^2}\right)^2
\Box^2\right]\phi_0,\\
\Pi_2&=&{\delta L\over\delta\partial_0^3\phi}
=c_4\left({(2\pi)^2\over\Lambda^2}\right)^2
\Box\phi_1.\nonumber
\eea
One can easily check that the Hamilton-Jacobi equations of 
the Hamiltonian
\bea\label{cham}
H&=&\int d^{d-1}x\left\{\sum\limits_{j=0}^2\Pi_j\partial_0\phi_j
-L(\phi,\partial\phi,\partial\partial\phi,\partial\partial\partial\phi)
\right\}\\
&=&\int d^{d-1}x\Biggl\{\Pi_0\phi_1+\Pi_1\phi_2
+{\Lambda^4\over2(2\pi)^4c_4}\Pi^2_2+\Pi_2\nabla^2\phi_1\nonu
&&-\hf(\phi_1^2+\phi_0\nabla^2\phi_0)
-c_2{(2\pi)^2\over2\Lambda^2}(\phi_2-\nabla^2\phi_0)^2\nonu
&&-{c_4\over2}\left({(2\pi)^2\over\Lambda^2}\right)^2
(\phi_2-\nabla^2\phi_0)\nabla^2(\phi_2-\nabla^2\phi_0)+V(\phi_0)
\Biggr\}\nonumber
\eea
are equivalent with the Euler-Lagrange equation. 

The quantization procedure is based on the 
canonical commutation relations
\be
\delta(x^0-y^0)[\phi_j(x),\Pi_k(y)]=i\delta_{j,k}\delta(x-y).
\ee
The inner consistency requires that $\phi_1$ and $\Pi_1$ be 
antihermitean operators. The standard path integral representation
\cite{bugr} results
\be
\langle\phi'_0,-\phi'_1,\phi'_2|e^{-iTH}|\phi_0,\phi_1,\phi_2\rangle
=\int D[\phi]D[\Pi]
e^{i\int_0^Tdt\{\sum_j\Pi_j\partial_0\phi_j-H[\Pi,\phi]\}},
\label{phpi}
\ee
where the fields $\Pi_1(x)$ and $\phi_1(x)$, as the eigenvalues of 
antihermitean operators are purely imaginary. This feature, the 
negative norm nature of the particle corresponding to the 
auxiliary field $\phi_1=\partial_0\phi$ makes the real time
phase space path integral \eq{phpi} divergent and ill defined
because the Hamiltonian \eq{cham} is complex,
\bea\label{conjug}
H[\Pi_0,\Pi_1,\Pi_2,\phi_0,\phi_1,\phi_2]^*&=&
H[\Pi_0,-\Pi_1,\Pi_2,\phi_0,-\phi_1,\phi_2]\nonu
&\not=&
H[\Pi_0,\Pi_1,\Pi_2,\phi_0,\phi_1,\phi_2].
\eea

In view of these failures of rescuing the model by the introduction 
of the auxiliary fields it seems remarkable that the Wick rotation
based on our lattice regularized Euclidean model does yield an 
acceptable theory when the elementary fields are chosen with more care. 
We shall show that the cutoff theory is unitary and stable
in the Minkowski space-time. The Lorentz symmetry is recovered
in the renormalization process only, when the cutoff is removed.

We start by noting that the elimination of 
a heavy particle which gave rise the
higher order derivative terms in the effective action 
actually creates mixed states from the pure ones. Thus the
effective theory should be recasted in terms
of its density matrix. This is not what happens in
the usual blocking procedure which
yields the path integral expression as a representation
of the matrix elements of the time evolution operator between 
pure field eigenstates. But the initial and final field 
configurations of the path integral
for the effective theory specify the states of the light particles
only. The information concerning the states of the heavy degrees
of freedom is lost in the elimination process, by the 
trace operation in their Hilbert space. In other words,
the initial and final states of the effective model specified 
in its path integral are mixed. The additional correlations in the expectation
values arising from the off-diagonal elements of the density
matrix are represented by the higher order derivative terms
of the effective action. 

Suppose that we truncate the effective
action up to $O(\partial^{2n})$. The correlations, i.e.
the information about the heavy particles in the
effective theory extend up to $n$ lattice spacing. The states
appearing in the path integrals can formally be considered as
pure ones if these correlations are properly kept in them.
This can be achieved by the introduction of $n-1$ auxiliary
fields at each site, like with the Pauli-Villars fields \eq{pavi}
or the canonical quantization \eq{phpi}. Another possibility
is the regroupment of $n$ consecutive lattice sites in 
the time direction with their field 
variables into a single $n$-component field\footnote{
The need of the increase of the number of degrees of 
freedom can be understood in the semiclassical limit, as well. 
The unique characterization of the trajectories interpolating 
between the initial and the final field configuration requires
the knowledge of $2n$ quantities.}, in a manner similar to the one
employed in ref. \cite{trmat}. We follow the
latter strategy, the reminiscent of the spin-flavor assignment
for the staggered fermions in real-space time \cite{stag}.
The corresponding action for 
\be
\phi_j(t,\vec x)=\phi(nt+j,\vec x), 
\ee
where $t$ is integer and $j=1,\cdots,n$ is
\be
S=\sum\limits_{t,\vec x,\vec y}\phi_j(t+1,\vec x)
\Delta^{-1}_{j,k}(\vec x,\vec y)\phi_k(t,\vec y)
+\sum\limits_{t,\vec x}\tilde V(\phi_j(t,\vec x)),
\label{blact}
\ee
where the space-time vectors $x=(t,\vec x)$ 
label the 'fat' lattice sites with $n$ degrees of freedom.

\begin{figure}
\centerline{\psfig{file=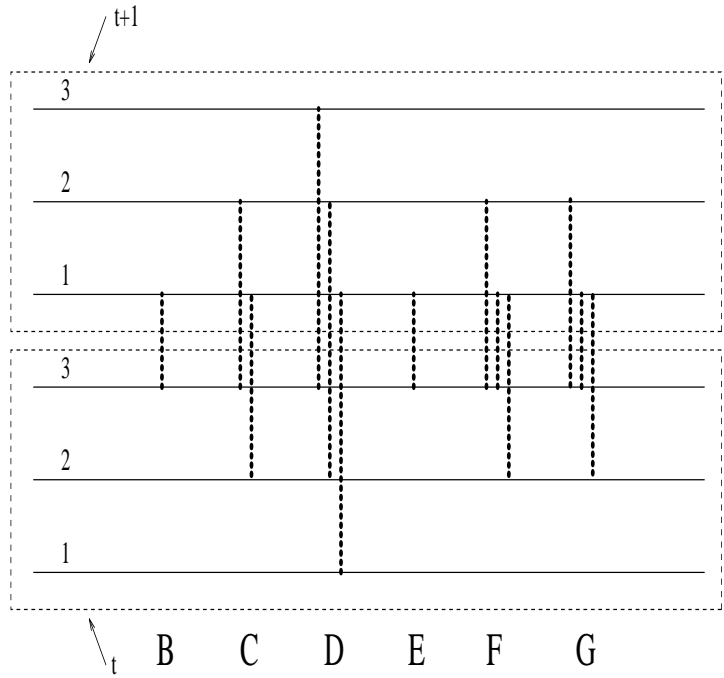,height=10cm,width=10cm,angle=270}}
\caption{\label{der} The coupling of the original time slices, indicated
by solid lines, between two 'fat' slices which are represented by the
boxes. The vertical dotted lines are the couplings, the nonzero elements
of $\Delta^{-1}_{j,k}$. The coefficients $B,~C,\cdots$ of the 
couplings are given in the lowest line.}
\end{figure}

In the case of the action \eq{res} we regroup three time slices
into one new 'fat' time slice $(n=3)$, $\phi_j(t,\vec x)=\phi(3t+j,\vec x)$
for $j=1,2,3$ and the matrix $\Delta^{-1}_{j,k}$ connects the
$(j,k)$ pairs shown in Fig. \ref{der},
\bea
\Delta^{-1}_{j,k}(\vec x,\vec y)&=&
2B\delta_{\vec x,\vec y}\delta_{j,1}\delta_{k,3}+
2C\delta_{\vec x,\vec y}(\delta_{j,2}\delta_{k,3}+\delta_{j,1}\delta_{k,2})
+2D\delta_{\vec x,\vec y}\delta_{j,k}\nonu
&+&\sum_j\biggl[2E(\delta_{\vec x,\vec y+e_j}+\delta_{\vec x,\vec y-e_j})
\delta_{j,1}\delta_{k,3}
+2F\biggl((\delta_{\vec x,\vec y+2e_j}+\delta_{\vec x,\vec y-2e_j})
\delta_{j,1}\delta_{k,3}\nonu
&&+(\delta_{\vec x,\vec y+e_j}+\delta_{\vec x,\vec y-e_j})
(\delta_{j,2}\delta_{k,3}+\delta_{j,1}\delta_{k,2})\biggr)\biggr]
\eea
in two dimensions $(F=0)$.
The transfer matrix for the field $\phi_j$ is given by
\be
T=e^{-3aH}
\ee
in terms of the original Hamiltonian.

Our reasoning is the following.
The unitarity of the time evolution operator for the states
created by the field operator $\phi(x)$ will be the result of the
realness of the spectrum of the square of the transfer matrix
with $n=3$, $T^2=e^{-6aH}$. The positivity of the spectrum
of $T^2$ and the positive definiteness of the norm in the
Hilbert space will be proven by verifying that the reflection 
positivity holds for $T$ \cite{ossc}. The cancellation
of the runaway amplitudes constructed perturbatively in
ref. \cite{lewi} is assured by the reflection positivity
for $T$ in a non-perturbative manner. The stability
of the vacuum, i.e. the boundedness of the spectrum of the 
Hamiltonian from below follows from the finiteness of the 
regulated Euclidean path integral. 

The proof of the reflection positivity relies on the Euclidean time 
inversion operator $\Theta$ defined as
\be
\Theta(t,\vec x)=(-t,\vec x),~~~\Theta[\phi_j(x)]=\phi_{4-j}(\Theta x),
\ee
where the transformation $j\to 4-j$ performs the time inversion
for the degrees of freedom within a 'fat' lattice site. 
Note that the base point of the inversion on the original lattice
for a given $n$ is at a lattice site or halfway between
two such sites for $n$ odd or even, respectively.
The time inversion $\Theta$ acts in the external, space-time coordinates
and the internal space of $j$ or other indices, it changes the sign of the time
coordinate and performs the necessary inversions in the
internal space, such as the transformation $j\to4-j$,
the complex conjugation in the case of complex quantities, respectively.
It is useful to introduce the internal time parity $T_F=\pm1$ 
as the eigenvalue of the time inversion in the internal space 
of an observable $F$ satisfying the condition
\be
\Theta\left[F[\phi(x)]\right]=T_FF\left[\phi(\Theta x)\right].
\ee

The reflection positivity of the transfer matrix $T$ 
can conveniently be expressed by means of the time inversion $\Theta$,
it is the requirement that the inequality
\be
\langle F\Theta[F]\rangle\ge0
\label{reflpo}
\ee
hold for any local functional $F[\phi]$ of the field variable
taken for $t>0$. The time evolution is clearly unitary in the subspace
of the Fock space which is span by the states obtained by the application
of the operator $F[\phi]$ on the vacuum. 

To prove \eq{reflpo} we write the action 
\eq{blact} as
\be
S=S_0+S_-+S_+
\label{acslp}
\ee
where
\bea
S_0&=&\sum_{\vec x}\tilde V(\phi_j(0,\vec x)),\nonu
S_-&=&\sum\limits_{t<0,\vec x,\vec y}\phi_j(t+1,\vec x)
\Delta^{-1}_{j,k}(\vec x,\vec y)\phi_k(t,\vec y)
+\sum\limits_{t<0,\vec x}\tilde V(\phi_j(t,\vec x)),\\
S_+&=&\sum\limits_{t\ge0,\vec x,\vec y}\phi_j(t+1,\vec x)
\Delta^{-1}_{j,k}(\vec x,\vec y)\phi_k(t,\vec y)
+\sum\limits_{t>0,\vec x}\tilde V(\phi_j(t,\vec x)).\nonumber
\eea
We shall need below the relation 
\be\label{acrp}
\Theta[S_\pm[\phi]]=S_\mp[\phi],
\ee
a consequence of the time reversal invariance. In terms of the matrix
$\Delta^{-1}_{j,k}(\vec x,\vec y)$ this amounts to the equations
\bea
\Theta[\sum_{t\ge0,\vec x,\vec y}\phi_j(t+1,\vec x)
\Delta^{-1}_{j,k}(\vec x,\vec y)\phi_k(t,\vec x)]
&=&\sum_{t\ge0,\vec x,\vec y}\Theta[\phi_j(t+1,\vec x)]
\Theta[\Delta^{-1}_{j,k}(\vec x,\vec y)]\Theta[\phi_k(t,\vec x)]\nonu
&=&\sum_{t\ge0,\vec x,\vec y}\phi_{4-j}(-t-1,\vec x)
\Theta[\Delta^{-1}_{j,k}(\vec x,\vec y)]\phi_{4-k}(-t,\vec x)]\nonu
&=&\sum_{t<0,\vec x,\vec y}\phi_{4-k}(t+1,\vec x)
\Theta[\Delta^{-1}_{j,k}(\vec x,\vec y)]\phi_{4-j}(t,\vec x)]\nonu
&=&\sum_{t<0,\vec x,\vec y}\phi_k(t+1,\vec x)
\Theta[\Delta^{-1}_{4-j,4-k}(\vec x,\vec y)]\phi_j(t,\vec x)]\nonu
&=&\sum_{t<0,\vec x,\vec y}\phi_k(t+1,\vec x)
\Delta^{-1}_{k,j}(\vec x,\vec y)\phi_j(t,\vec x)
\eea
which yield immediately
\be\label{tinip}
\Theta[\Delta^{-1}_{j,k}(\vec x,\vec y)]
=\Delta^{-1}_{4-k,4-j}(\vec x,\vec y)
=\Delta^{-1}_{j,k}(\vec x,\vec y).
\ee

The left hand side of the inequality \eq{reflpo} can now be written as
\bea\label{prrp}
\langle F\Theta F\rangle&=&
\int D_{t=0}[\phi]e^{-S_0[\phi(x)]}
\int D_{t<0}[\phi]e^{-S_-[\phi(x)]}\Theta\left[F[\phi(x)]\right]
\int D_{t>0}[\phi]e^{-S_+[\phi(x)]}F[\phi(x)]\nonu
&=&\int D[\phi]e^{-S_0[\phi(x)]}\Theta\left[e^{-S_+[\phi(x)]}\right]
\Theta\left[F[\phi(x)]\right]e^{-S_+[\phi(x)]}F[\phi(x)]\nonu
&=&\int D[\phi]e^{-S_0[\phi(x)]}e^{-S_+[\phi(\Theta x)]}
T_FF\left[\phi(\Theta x)\right]e^{-S_+[\phi(x)]}F[\phi(x)],
\eea
where we assumed in the last equation that the functional $F$ possesses a definite internal
time inversion parity, $T_F=\pm1$. This allows us to
eliminate $\Theta$ from the integrand by writing
\bea\label{prrpp}
\langle F\Theta F\rangle=\int D_{t=0}[\phi]e^{-S_0}T_F
\left(\int D_{t>0}[\phi]e^{-S_+}F[\phi(x)]\right)^2.
\eea
The result is the inequality
\be
T_F\langle F\Theta[F]\rangle\ge0,
\label{reflpog}
\ee
holding for any real functional $F$. 

Let us denote by ${\cal H}_1$ the subspace of the Fock space $\cal H$
which is span by the application of the $+1$ internal 
time parity local functionals of the field operator $\phi(x)$ 
on the vacuum and by ${\cal H}_{orth}$ its orthogonal complement. 
We have 
\be
{\cal H}={\cal H}_1\oplus{\cal H}_{orth}
\ee
in an obvious manner. The time evolution is unitary within
${\cal H}_1$, and the vectors in ${\cal H}_{orth}$, among others 
the ghost states of ref. \cite{hiderk}, have negative norm in 
agreement with \eq{conjug}.

More care is needed in the presence of a time dependent condensate. 
This is because the condensed mode is treated classically 
and the problems of the quantum treatement, in particular the 
unitarity of the time evolution and the reflection 
positivity apply for the quantum fluctuations only.
The reflection positivity should hold for the 
dynamics of the quantum variable $\phi_q(x)=\phi(x)-\la\phi(x)\ra$
which is governed by the action 
\be
S_q[\phi_q]=S[\phi_q+\la\phi\ra].
\ee
Note that \eq{acrp} is satisfied by $S_q$,
\be\label{acrpg}
\Theta[S_\pm[\phi_q+\la\phi\ra]]=S_\mp[\phi_q+\la\phi\ra],
\ee
according to \eq{tinip}. This relation allows us to repeat
the steps \eq{prrp}-\eq{prrpp} after having performed
the replacement $S[\phi]\to S_q[\phi]$. 

Another problem one
has to take care is the choice of $n$, the number of degrees of
freedom at a 'fat' lattice site. $n$ must be at least half of the
highest power the gradient operator appears in the action in order 
to decouple the next-to-nearest neighbors. In the same time it
must be integer times the period length of the elementary cell in the
antiferromagnetic phase. These conditions require the choice
$n=4$ when $N=2$ and $n=N$ for $N\ge3$. 

The reflection positivity
satisfied in the subspace ${\cal H}_1$ of the $N=2$ antiferromagnetic
phase indicates that one has to use the space parity eigenstates, the
superpositions of the two chiral bosons observed at $c_2=1/3d$,
$c_4=0$ \cite{ehj}, because only the scalar particle is physical, the 
pseudoscalar modes correspond to ghost states.

\section{Conclusions}
A simple two dimensional effective theory was considered in this work for an
elementary interaction which is strongly repulsive for short distances and 
attractive at long distances and supposed to form a crystalline ground state. 
Such an interaction is 
coded in terms of the effective action with higher order of the derivatives. 
The inhomogeneous ground state is reproduced in the saddle point
approximation of the effective theory. 

The mean-field, tree level phase structure of the
model is quite involved and displays several inhomogeneous phases
with a number of commensurate-incommensurate transitions. 
The elementary excitations contain modes similar to the
rotons of liquid $He^4$ but the excitation spectrum is usually
massive in lacking of a conserved particle number. 
Our effective theory is a simplified version of the higher dimensional
systems with inhomogeneous ground states, such as the solids,
antiferromagnets and materials with charge density phase. The 
formation of the inhomogeneous vacuum by the condensation of
particles with non-vanishing momentum is a manifestation of 
the additional soft modes which characterizes the quantum 
phase transtitions. 

The one-loop corrections give a line of ultraviolet fixed points with
variable particle content. The period length of the vacuum can thus be
sent to zero. The Poincare symmetry is restored in the continuum limit
and the vacuum becomes homogeneous for the measurements made at finite energies. 
In the same time the excitation spectrum of the model 
remains nontrivial, reflecting the inhomogeneity of the vacuum. 
The different dispersion relation branches are interpreted in the continuum
limit as excitations with different flavor. This unusual vacuum-excitations
correspondence opens the way for the construction of new kind of unified
quantum field theoretical models where several particles are described by the same field.
Our effective theory was found to have a consistent extension to real time
in the even internal time parity sector of the Fock space
despite the presence of the higher order derivative terms in the action.

The numerical analysis performed by the Monte Carlo method confirms the
mean-field prediction of the phase structure and is consistent 
with the criticality at the
ferromagnetic phase transition line. The frustrations are the slow modes
of the simulation when the amplitude of the modulated vacuum is large
suggesting the possibility of a quenched disorder
variable in real time, as well. 

We note finally that there are no massless excitation modes above the modulated 
vacuum with finite period length in lattice spacing units. This is specially 
striking in two dimensions where the periodic structure of the periodic 
vacuum is not necessarily destroyed by the Mermin-Wagner-Coleman theorem \cite{mwc}.
The continuum limit of the antiferromagnetic phase where $N/M\to\infty$ 
might be similar to the planar X-Y model with power like decay 
of the correlations.

\section{Acknowledgement}
We thank Vincenzo Branchina for useful discussions 
and the referee for his constructive critics.
This work was supported in part by the grant OTKA T29927/98
of the Hungarian Academy of Sciences.

\begin{table}
\begin{center}
\begin{tabular}{@{}*{3}{|l||l|l}}
&$d_{AF}=1$&$d_{AF}=2$\cr
\hline    
\hline
$N=2$  &$\left(1+{4(1-4c_2+16c_4)\over m^2}\right)^2$
       &$\left(1+{8(1-8c_2+64c_4)\over m^2}\right)^2$\cr
$N=3$  &${2\over3}\left(1+{3(1-3c_2+9c_4)\over m^2}\right)^2$
       &${4\over9}\left(1+{6(1-6c_2+36c_4)\over m^2}\right)^2$\cr
$N=4$  &$\left(1+{2(1-2c_2}\right)^2$
       &$\left(1+{4(1-4c_2+16c_4)\over m^2}\right)^2$\cr
$N>4$  &${2\over3}\left(1+{4s^2(1-4s^2c_2+16s^4c_4)\over m^2}\right)^2$
       &${4\over9}\left(1+{8s^2(1-8s^2c_2+64s^4c_4)\over m^2}\right)^2$\cr
\end{tabular}
\caption{The antiferromagnetic action density, $s(d_{AF},N)/s(0,N)$. 
The phase angle has been chosen to be $\alpha=\pi/4$ for $N=4$ to 
minimize the action and $s=\sin\pi/N$, $N>4$. Since $s(0,N)<0$ the
antiferromagnetic phase is preferred against the ferromagnetic one
whenever the corresponding expression in the table is larger than 1. 
\label{esperes}}
\end{center}
\end{table}

\end{document}